\documentclass[%
  reprint,
  amsmath,amssymb,
  floatfix,
  ]{revtex4-2}

  \usepackage{graphicx}% Include figure files
  \usepackage{dcolumn}% Align table columns on decimal point
  \usepackage{bm}% bold math
  \usepackage{comment}
  \usepackage{graphicx}
  \usepackage{ascmac}
  \usepackage{mathtools}
  \usepackage{color}
  \usepackage{xcolor}
  \usepackage{overpic}
  \usepackage[whole]{bxcjkjatype}

\begin{document}

  \preprint{APS/123-QED}
  \title{
    Reciprocal microswimming in fluctuating and confined environments
  }

  \author{Yoshiki Hiruta}
  \email{hiruta@kurims.kyoto-u.ac.jp}
  \author{Kenta Ishimoto}
  \email{ishimoto@kurims.kyoto-u.ac.jp}
  \affiliation{
    Research Institute for Mathematical Sciences, Kyoto University, Kyoto 606-8502, Japan}
  \date{\today}% It is always \today, today,

  \date{\today}

  \begin{abstract}
  From bacteria and sperm cells to artificial microrobots, self-propelled microscopic objects at low Reynolds numbers often perceive fluctuating mechanical and chemical stimuli and contact exterior wall boundaries both in nature and the laboratory.
  In this study, we theoretically investigate the fundamental features of microswimmers  by focusing on their reciprocal deformation. Although the scallop theorem prohibits the net locomotion of reciprocal microswimmers, by analyzing a two-sphere swimmer model, we show that in a fluctuating and geometrically confined environment, reciprocal deformations can afford a statistically average displacement. After designing the shape gait, a reciprocal swimmer can migrate in any direction, even in the statistical sense, while the statistical average of passive rigid particles statistically diffuses in a particular direction in the presence of external boundaries.
  To elucidate this symmetry breakdown, by introducing an impulse response function, we derive a general formula for predicting the nonzero net displacement of a reciprocal swimmer.
  Using this theory, we determine the relation between the shape gait and net locomotion as well as the net diffusion constant increase and decrease owing to a reciprocal deformation.
  Based on these findings and a theoretical formulation, we provide a fundamental basis for environment-coupled statistical locomotion. 
  Thus, this study is valuable for understanding biophysical phenomena in fluctuating environments, designing artificial microrobots, and conducting laboratory experiments.
  \end{abstract}

  \maketitle

  \section{Introduction}

Even one water droplet in a pond contains thousands of swimming cells with diverse morphologies . These cells respond to physical and chemical stimuli from external environments; their response behavior can be active or passive. Active responses are well known in chemotaxis, for example, sperm cells responding to molecules released from eggs or bacteria swimming toward nutrients and oxygen \cite{friedrich2007chemotaxis, shiba2008ca2+, desai2019hydrodynamic, hillesdon1996bioconvection}. 
Some microorganisms and cells turn their heads upstream, which is considered a passive behavior without sensing external stimuli
because of hydrodynamic interactions between the self-propelled object and external fluid flow \cite{kaya2012direct, kantsler2014rheotaxis, ishimoto2015fluid, mathijssen2019oscillatory, ishimoto2023jeffery}.
Microscopic, self-propelled objects are called microswimmers, which have served as inspiration for the study of artificial active agents in recent years \cite{soto2022smart, essafri2022designing}. Further, enabled by the rapid development of machine learning, a new research field has emerged that focuses on creating and designing smart particles that can change their dynamics by perceiving external environments and learning from them \cite{tsang2020roads, stark2021artificial}.

Notably, the natural environment fluctuates over time. In particular, the mentioned microscopic objects are frequently subjected to Brownian thermal noise.
Another important physical feature of environments is external boundaries, such as the air--water surface and wall substrates. Microswimmers are also known to accumulate near walls purely via physical mechanisms, such as hydrodynamic and contact interactions \cite{takagi2014hydrodynamic, ishimoto2014study, shum2015hydrodynamic, chamolly2017active}.
Previous studies have focused on microswimming inside spherical rigid containers and liquid droplets \cite{reigh2017,shaik2018,huang2020,sprenger2020}.
In this study, we consider the general aspect of microswimming in noisy and confined environments.

Owing to the small size of microswimmers, inertial effects are negligible; thus, the flow around a swimmer is governed by the Stokes equation describing the dynamics of low-Reynolds-number flows.
Because of the time-reversal symmetry associated with the Stokes flow, a reciprocal deformation cannot result in net locomotion; this is known as the scallop theorem \cite{Purcell1977, Shapere1989, Lauga2011, ishimoto2012coordinate}.
Therefore, one degree of freedom is insufficient for propulsion. An example is a two-sphere model, which comprises two spheres connected by a rod. Although a swimmer can vary the rod length, only reciprocal deformation is possible. Thus, it remains motionless in a Newtonian fluid \cite{dunstan2012, datt2018two, eberhard2023}.
Therefore, microswimmers frequently utilize nonreciprocal deformations, such as multiple degrees of freedom, and this behavior can be well studied using a three-sphere model \cite{najafi2004simple, golestanian2008analytic, yasuda2023generalized}.
The scallop theorem only holds when the fluid equation satisfies the Stokes equation and the swimmer inertia is negligible. Finite inertia effects, therefore, result in net locomotion even with reciprocal deformations \cite{lauga2007continuous, ishimoto2013spherical, hubert2021scallop, derr2022reciprocal}.
Additionally, non-Newtonian fluids, such as viscoelastic fluids, allow reciprocal swimmers to generate net locomotion \cite{Curtis2013, lauga2014locomotion, qiu2014swimming, datt2018two, yasuda2020reciprocal}.
However, external boundaries do not break the scallop theorem; the theorem still holds if the Stokes flow is satisfied.

The effects of thermal noise on an active swimmer have been intensively studied using nondeforming active agents, such as active Brownian particles \cite{romanczuk2012active, Bechinger2016, patch2017kinetics,bar2020self,yasuda2022most}.
Furthermore, the swimming problem with fluctuating shape gaits has been well studied, for example, using three-sphere models \cite{golestanian2008mechanical, yasuda2021odd}. In this case, the swimming velocity is determined based on the statistical average of an area enclosed in the shape space, and the reciprocal swimmer does not generate net locomotion \cite{ishimoto2022self, ishimoto2023odd}. This is, however, distinct from the problem where a swimmer's position fluctuates with fluctuations in the environment.

If the noise is spatially homogeneous, such as Brownian motion in free space, the statistical average of particle position does not change. However, in case of an external boundary, the environmental noise is no longer spatially homogeneous because of position-dependent hydrodynamic resistance \cite{matse2017test}. 
Particle diffusion is then suppressed near the boundary, and the averaged position moves away from it, with the averaged position evolving in time as $\mathcal{O}(\sqrt{t})$.
In contrast, studies on swimmers with deformations in noisy environments are still limited.
Hosaka et al. \cite{hosaka2017thermally} considered a three-sphere model with elastic springs. Each sphere possessed different temperatures and thus different noise magnitudes. 
They also theoretically demonstrated that the two-sphere model cannot generate net locomotion in a statistical sense in free space. However, the motions of a reciprocal swimmer under spatially inhomogeneous noise, such as geometrical confinement, are still unclear. 

Therefore, the primary aim of this study is to examine the effects of environmental noise on reciprocal swimmers, particularly in a geometrically confined environment. By considering a two-sphere swimmer model , we numerically and theoretically demonstrate that a reciprocal swimmer can generate net locomotion in a statistical sense. The key mathematical structure behind this result lies in the fact that the statistical average of the distribution function of a variable is, in general, different from the distribution function of the averaged variable.
To analyze a precise effect from environmental noise, we derive a statistical theory with low-order moments in the probability distribution function. Accordingly, the secondary aim of this study is to apply this theory to examine the effects of shape gaits on a swimmer's net velocity and net diffusion. In particular, we show that a reciprocal swimmer moves by linearly increasing its position with time (i.e., $\sim O(t)$). Notably, this active displacement dominates the displacement of the averaged position purely through passive diffusion ($\sim O(\sqrt{t})$) after a long time. Further, based on this analysis, we demonstrate that by choosing the shape gait of the deformation, a reciprocal swimmer can migrate in an arbitrary direction, which is distinct from the unidirection diffusive transport of a passive particle near an external boundary.

The remainder of this paper is organized as follows. In Sec. \ref{sec:setting}, we introduce the governing equation of a two-sphere swimmer model with an external boundary. Considering a small amplitude and far-field asymptotic regime, we provide an explicit form of the stochastic equation of the swimmer in a noisy environment. In Sec. \ref{sec:fundmental}, we provide numerical results to demonstrate that reciprocal deformations lead to net locomotion.
In Sec. \ref{sec:rtheory}, we present a theory to predict nonzero displacements, and in Sec. \ref{sec:VelocAndDiff}, we apply this theory to understand the symmetric properties of the shape gait and its impact on the net velocity and diffusion of the swimmer. Concluding remarks are provided in Sec. \ref{sec:conc}.

  \section{\label{sec:setting}Model microswimmers in a noisy and confined domain}

  \subsection{\label{sec:derivation} Equations of motion for a two-sphere swimmer}
  In this section, we introduce the equations of motion for a two-sphere swimmer under geometrical confinement. For simplicity, we assume that the swimmer's position is restricted to one dimension and neglect rotational motion.

\begin{figure}
    \centering
      \begin{overpic}
       [width=0.99\linewidth]{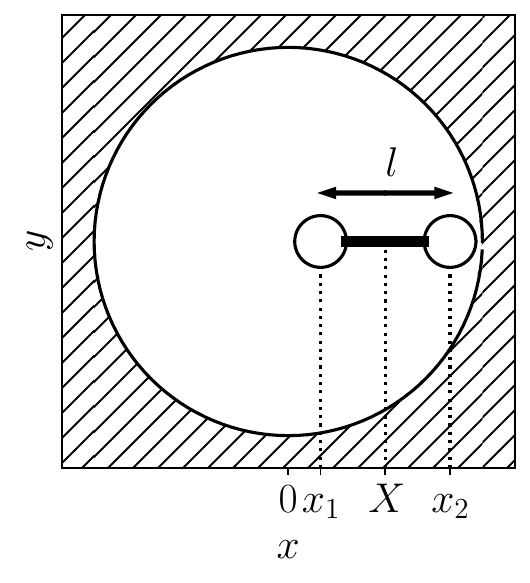}
      \put(0,95){{\large (a)}}
      \end{overpic}\\
      \begin{overpic}
    [width=0.49\linewidth]{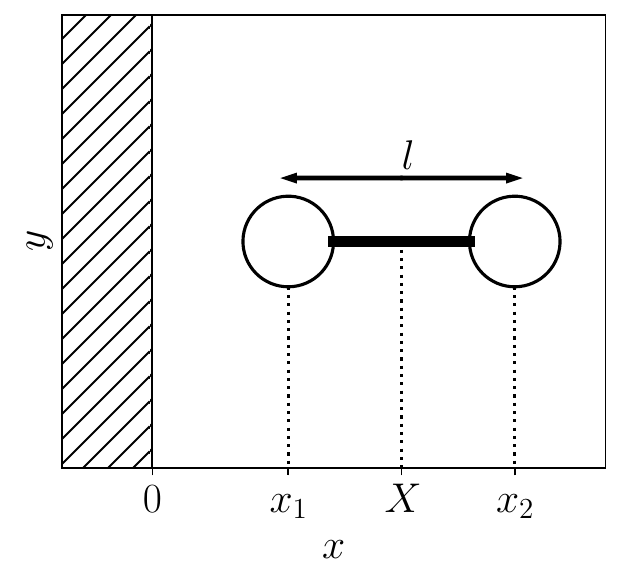}
      \put(-5,90){{\large (b)}}
      \end{overpic}
      \begin{overpic}
       [width=0.43\linewidth]{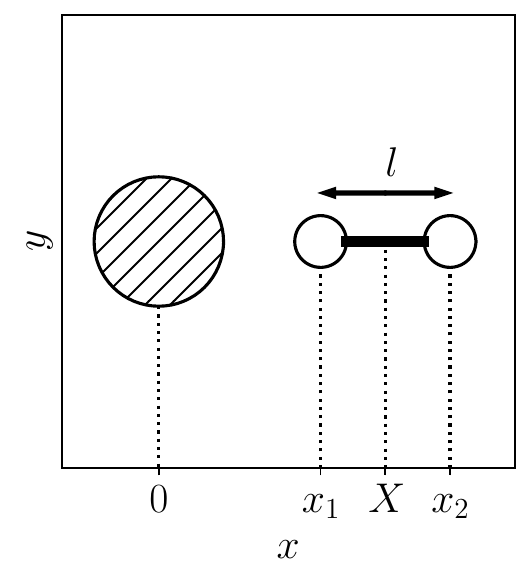}
      \put(-3,90){{\large (c)}}
      \end{overpic}
    \caption{(a) Schematic of the two-sphere swimmer inside a sphere. The two spheres with a radius of $a$ are connected through a rod of length $l(t)=x_2-x_1$, which designates the shape gait. The position of the swimmer, denoted by $X$, moves along the $x$ axis.
  The shaded area indicates a material wall. Schematics of the two-sphere swimmer near (b) a flat wall and (c) spherical obstacle.
  }
    \label{fig:swimmer}
  \end{figure}

  The two-sphere swimmer comprises two spheres of radius $a$ connected through a rod whose length is controlled as a function of time. Let $x_1$ and $x_2 > x_1$ be the center positions of the spheres. We define the swimmer position by $X=(x_1+x_2)/2$. Thus, the relative distance between the spheres, $l=x_2-x_1$, represents the swimmer configuration. Subsequently, we calculate the swimmer velocity $U=dX/dt$ for a given deformation $l(t)$ (FIG.\ref{fig:swimmer}). The function $l(t)$ designates the shape gait, and we consider a time-periodic deformation with a period $T$.

  We assume that the surrounding fluid obeys the Stokes equation; that is, for the velocity field $\bm{u}$ and pressure $p$,
 \begin{align}
    \bm{\nabla} p&=\mu\nabla^2\bm{u}\label{eq:stokes1}\\
    \bm{\nabla}\cdot \bm{u}&=0,\label{eq:stokes2}
 \end{align}
where constant $\mu$ represents the (dynamic) viscosity.
Owing to to the linearity of the Stokes flow, the hydrodynamic force on the swimmer can be decomposed into drag $F_{\textrm{d}}$ and propulsion $F_{\textrm{p}}$ \cite{lauga2009hydrodynamics, yariv2006self, ishimoto2012coordinate}. The drag $F_{\textrm{d}}$ is proportional to the velocity $dX/dt=U(X,l)$ and is thus written as $F_{\textrm{d}}=-M_{\textrm{d}}U$, where $M_{\textrm{d}}(X, l)$ indicates the drag coefficient and is positive-definite. Similarly, the propulsion is proportional to the rate of deformation, $F_{\textrm{p}}=M_{\textrm{p}}(dl/dt)$, with the coefficient $M_{\textrm{p}}(X, l)$ being a function of $X$ and $l$.
  Because of the negligible inertia of the swimmer, the total force on the swimmer is balanced; therefore, we can derive the one-dimensional equations of the swimmer without noise via \cite{lauga2009hydrodynamics}:
    \begin{align}
      U(X,l)&=M(X,l)\frac{dl}{dt}
      \label{eq:EOMdet},
    \end{align}
where the function $M(X,l)$ encodes hydrodynamic interactions, which are only determined by the instantaneous configuration of the swimmer and the geometry of the surrounding objects. Notably, Eq. \eqref{eq:EOMdet} holds for a general reciprocal swimmer moving in one direction once the function $l(t)$ reads any function to specify its shape gait. Moreover, most of the theoretical results presented in the following sections are not restricted to a specific model.
Given a shape gait $l(t)$ and an initial position $X(0)$, we can obtain the solution of Eq. \eqref{eq:EOMdet} as $X=X_0(t)$, which we hereafter call the {\it zero-noise solution}.

Under a noisy environment, the swimmer position $X$ becomes stochastic, and we denote its probability distribution function by $P(X,t)$.
A statistical spatial average of an arbitrary function $A(X)$ is defined as $\langle A \rangle_t\equiv\int dX A(X)P(X,t)$.
    We then introduce the average position $\langle X\rangle_t$ and variance $\langle X^2\rangle_t$, both of which are functions of time.

In addition to drag and propulsion, an external force $F_{\rm ext}(t)$ acts on the swimmer through the environmental noise. These are balanced as $F_{\rm d}+F_{\rm p}+F_{\rm ext}=0$ owing to the negligible inertia. Next, we introduce the Gaussian white noise $\xi(t)$ as an external force from the environment.

Owing to the position-dependent drag coefficient $M(X, l)$, the equations of motions generally contain a multiplicative noise term \cite{Sancho1982}, resulting in a position-dependent diffusion coefficient. Such a stochastic differential equation is analyzed as a Stratonovich type, producing an additional virtual-background flow term. This effective drift leads to nonzero net displacement for a reciprocal swimmer, even with noise. 

To simplify our analysis, we consider an asymptotic regime of $a\ll l \ll L_W$, where $L_W$ is the distance between the swimmer and wall boundary. Notably, the position dependence of the diffusion constant is negligible, as detailed and discussed in Appendix \ref{app:derivEoM}. Thus, our governing equation of a swimmer with environmental noise is written as
    \begin{equation}
    \frac{dX}{dt}=M(X, l)\frac{dl}{dt}+\sqrt{D}\,\xi(t)
    \label{eq:EOM},
    \end{equation}
and the random variable $\xi(t)$ represents the zero-mean normal white Gaussian noise, where $\langle\xi(t)\rangle =0$ and $\langle\xi(t_1)\xi(t_2) \rangle =\delta(t_1-t_2)$. Here, $\delta(t)$ is the Dirac delta function and the constant $D$ denotes the diffusion coefficient, representing the noise strength.

    From thermodynamics constraints, the diffusion constant $D$ should relate to the drag coefficient $\gamma$ through the fluctuation-dissipation theorem, an example of which is the Einstein relation $D= 2k_B\Theta\gamma^{-1}$ for a thermal noise of temperature $\Theta$ and Boltzmann constant $k_B$. Notably, the prefactor of 2 in the Einstein relation reflects our definition of the diffusion constant in Eq. \eqref{eq:EOM}. We can estimate a typical value of $D$ for a micron-scale colloid in water at room temperature as $DT/a^2 \approx 0.5$, where $T$ is the time period of the deformation. We use $\Theta=300$K, $a=1\mu$m, and $T=1$s.

      The stochastic equations of motion (\ref{eq:EOM}) are equivalent to the following Fokker--Planck equation for the probabilistic distribution $P(X,t)$:
    \begin{align}
    \frac{\partial}{\partial t}P&=-\frac{\partial}{\partial X}\left(M(X,l)\frac{dl}{d t}P\right)+\frac{D}{2}\frac{\partial^2}{\partial X^2}P \label{eq:FP}.
    \end{align}
    By definition, the average over white Gaussian noise for the stochastic Langevin equation is equivalent to the average for the Fokker-Planck equation with respect to $P$.
  In the subsequent sections, we numerically solve Eq. (\ref{eq:FP}) using a finite-volume method with second-order spatial accuracy %for spatial direction  
 and the fourth Runge--Kutta method with space and 
  time discretizations of $0.01a$ and $10^{-5}T$, respectively. 
  We impose no-flux boundary conditions at the spatial boundary, which is taken at an adequate distance from the swimmer. 
  The initial condition is set to be a Gaussian distribution; that is, $P(X)\propto\exp(-(X-X(0))^2/\sigma_0^2)$, with the parameter $\sigma_0$ being taken as sufficiently small.

We now discuss that the averaged position is generally different from the position in the deterministic case (i.e., $\langle X\rangle_t\neq X_0(t)$). By taking an ensemble average in Eq. \eqref{eq:EOMdet}, we obtain
\begin{align}
      \langle U \rangle_t&=\langle M(X,l) \rangle_t \frac{dl}{dt}
      \label{eq:Uave}.
\end{align}
Hence, the equality $ \langle X \rangle_t=X_0(t)$ holds only if the relation $\langle M(X,l) \rangle = M(\langle X \rangle ,l)$ is satisfied at an arbitrary time $t$. This, however, does not generally hold for $M(X)$. In fact, using Jensen's inequality for a lower-convex function $M(X)$, such as $M=X^2$, we have $\langle M(X,l)\rangle \geq M(\langle X\rangle ,l).$ The equality only holds when $M$ is linear or a constant in $X$, and the leading-order effects of the convexity should therefore be captured by the second spatial derivative of $M$. Such spatial dependence of $M(X)$ is possible when the system does not exhibit translational symmetry, and external boundaries generally break this symmetry, implying that that the scallop theorem is violated in a statistical sense under a noisy and confined environment regardless of the details of the function of $M$.%; hence, detailed swimmer models are required .

Nonetheless, these simple arguments only consider the instantaneous velocity and do not exclude the possibility that the averaged net displacement $\langle X\rangle_T$ cancels out after one deformation cycle $T$, while the zero-noise solution $X_0(T)$ must vanish owing to the scallop theorem. We therefore assume particular geometries for the external boundaries to further quantify the net displacement under thermal environmental noise.

  \subsection{Swimmer--wall hydrodynamic interactions
  \label{sec:CalcM}
  }

Henceforth, we focus on the two-sphere swimmer inside a spherical container and with a spherical obstacle or flat wall, where the sphere center and wall are located at $x=0$ and position $X$ is assumed to be positive without loss of generality (FIG.\ref{fig:swimmer}). With these simple geometries, we derive the leading-order dependence of $M(X, l)$  on $X$.

  First, we estimate the force $F_\alpha$ acting on the sphere $\alpha$ $(\alpha \in\{1,2\})$ from the surrounding fluid using the Fax\'en law. With our asymptotic regime of $a/l \ll 1$, by dropping the $O((a/l)^2)$ correction, we can write the force using the local Stokes drag with a correction term as
  
    \begin{align}
      F_\alpha=-\gamma \left(\frac{dx_\alpha}{dt}-U^{I}_{\alpha}\right), \label{eq:Fi}
    \end{align}
where $\gamma=6\pi\mu a$ is the Stokes drag coefficient and $U^{I}_{\alpha}$ is the perturbed fluid velocity at position $x_\alpha$ 
owing to the motion of the other sphere and the presence of the external boundary.

In this asymptotic regime, where the effects of the sphere size are negligible, the perturbed velocity $U^{I}_{\alpha}$ can be written using Green's function of the Stokes flow with the boundary geometry, as follows:
    \begin{equation}
    U_{\alpha}^{I}=-\tilde{G}_{\alpha\beta}F_\beta\label{eq:UI}, 
    \end{equation}
where $\tilde{G}_{\alpha\beta}$ is the summation of the Stokeslet along the $x$ axis and corrections due to the presence of a wall boundary, $G^{\rm W}(x_\alpha,x_\beta)$.
Let us introduce the Stokeslet in three dimensions as $\mathcal{G}_{ij}(\bm{x}, \bm{0})=(8\pi\mu)^{-1}(r^{-1}+x_ix_jr^{-3})$ for a point $\bm{x}$ in three dimensions, where the indices $i$ and $j$ denote the spatial coordinates, $i, j\in \{1,2,3\}$, and $r=|\bm{x}|$. Now, all the dynamics are along the $x$ axis. By substituting $\bm{x}_\alpha=(x_\alpha, 0, 0)$ into the Stokeslet, we introduce sphere--sphere interactions using $G^{\textrm{S}}(x_\alpha, x_\beta)=\mathcal{G}_{11}(\bm{x}_\alpha, \bm{x}_\beta)$ for $\alpha\neq \beta$ and zero otherwise.
Namely,
    \begin{equation}
    G^{\rm S}(x_\alpha,x_\beta)=\frac{1}{4\pi\mu|x_\alpha-x_\beta|} {\rm ~~for~~} \alpha\neq \beta
    \label{eq:stokeslet}
    \end{equation}
and is zero when $\alpha=\beta$ as the self-induced velocity should be removed. Therefore, the Green's function in \eqref{eq:UI} is written as
    \begin{align}
      \tilde{G}_{\alpha\beta}= G^{\rm S}(x_\alpha,x_\beta)+G^{\rm W}(x_\alpha,x_\beta).
      \label{eq:Gtilde}
    \end{align}
  Notably, $\tilde{G}_{\alpha\beta}=\tilde{G}_{\beta\alpha}$ and is valid for any boundary shape \cite{Pozrikidis1992}.
  From Eqs. (\ref{eq:Fi}) and (\ref{eq:UI}), we can derive an equation to determine the force acting on each sphere as follows:
    \begin{align}
      F_\alpha&=-\left(\frac{I}{\gamma}+\tilde{G}\right)_{\alpha\beta}^{-1} \frac{dx_\beta}{dt} \label{eq:Feq},
    \end{align}
where $I$ is the identity matrix.
  Generally, the sum of the forces acting on the microswimmer, $F_1+F_2$, may be decomposed into the drag force $F_{\rm d}$, which is proportional to $U$, and $F_{\rm p}$, which is proportional to $dl/dt$, as
    \begin{align}
      F_{\rm d}&=\frac{-1}{{\rm det}(\gamma^{-1}I+\tilde{G})} \nonumber \\
      &\times 
      \left(2\gamma^{-1}+\tilde{G}_{11}+\tilde{G}_{22}-\tilde{G}_{12}-\tilde{G}_{21}\right)\frac{dX}{dt}\label{eq:Fd}\\
      F_{\rm p}&=\frac{-1}{{\rm det}(\gamma^{-1}I+\tilde{G})}\left(\tilde{G}_{11}-\tilde{G}_{22}\right)\frac{dl}{dt}\label{eq:Ft}.
    \end{align}
In the asymptotic regime of $a\ll l$, the leading-order expression is ${\rm det}(\gamma^{-1}I+\tilde{G})=\gamma^{-2}\left(1+O(a/l)\right)$, which yields the following asymptotic form: 
    \begin{align}
      F_{\rm d}&\simeq-2\gamma \frac{dX}{dt}\label{eq:Fd2}\\
    F_{\rm p}&\simeq-\gamma^2 \left(\tilde{G}_{11}-\tilde{G}_{22} \right)\frac{dl}{dt}\label{eq:Ft2},
    \end{align}
where the neglected error terms are of the order of $O(a/l)$.
The prefactor of 2 in Eq. \eqref{eq:Fd2} reflects the two spheres. The propulsion force represented in Eq. \eqref{eq:Ft2} includes the wall-induced correction, and we also confirm that no motion is generated without an external boundary because both $\tilde{G}_{11}$ and $\tilde{G}_{22}$ include only wall-induced flows.

    Therefore, in the equations of motion [Eq. \eqref{eq:EOM}], $M$ is written in the following form:
    \begin{align}
    &M(X,l)= \nonumber \\
    &\frac{\gamma}{2} \left[G^{\rm W}\left(X+\frac{l}{2},X+\frac{l}{2}\right)-G^{\rm W}\left(X-\frac{l}{2},X-\frac{l}{2}\right)\right]\label{eq:EOM2}.
    \end{align}

We then proceed to derive expressions for the $M(X,l)$ in Eq. \eqref{eq:EOM2} for a two-sphere swimmer in three geometries (FIG.\ref{fig:swimmer}): $M_{\rm int}$ for a swimmer inside a spherical container of radius $R$,
  $M_{\rm flat}$ for the presence of an infinite flat wall, and $M_{\rm ext}$ for the presence of an external spherical obstacle with a radius $R$.

  The form of the boundary-induced Green's function $G^{W\rm }(X, X)$ is explicitly solved using the method of images and obtained as the superpositions of a Stokeslet and its multipoles located in the region outside the fluid domain \cite{maul1994image, kim2013microhydrodynamics}. The expression is, therefore, written as
  \begin{equation}
   G^{\rm W}(X, X)=\alpha(X) G^{\rm S}(X-X_{\rm IM}(X)) 
   \label{eq:Gw},
  \end{equation}
  with the strength and position of the mirror image represented by $\alpha$ and $X_{\rm IM}$, respectively, both of which are functions of the position $X(>0)$. 
  Next, we derive the leading-order contribution of Green's function $G^{\rm W}(X, X)$ by 
   assuming that the swimmer is far enough away from the external boundaries.
  
 When a flat wall is located at $x=0$ [FIG. \ref{fig:swimmer}(b)], by evaluating the weights of the mirror singularities to impose the no-slip condition at the wall boundary, we obtain $\alpha=-1$ and $X_{\rm IM}=-X$. Thus, we have the leading-order contribution
  \begin{equation}
      G^{\rm W}(X,X)=-\frac{1}{4\pi\mu|X-{X_{\rm IM}}|}=-\frac{1}{8\pi\mu X}, \label{eq:Gflat}
  \end{equation}
  and by substituting Eq. \eqref{eq:Gflat} into Eq. \eqref{eq:EOM2}, we obtain
  \begin{equation}
  M_{\rm flat}  
  =\frac{-\gamma}{16\pi\mu}\frac{l}{X^2}  
  \label{eq:Mwall}.
  \end{equation} 

  When the swimmer is situated exterior or interior to a sphere [FIGs. \ref{fig:swimmer}(a, c)], a similar analysis of the weights of the mirror singularities provides the parameters $\alpha=-3R/2X$
 and $X_{\rm IM}=R^2/X$. Thus, we obtain a similar asymptotic form of $G^{\rm W}$:%$M$
  \begin{equation}
   G^{\rm W}(X,X)=\frac{\pm 3R}{8\pi\mu (X^2-R^2)},
  \end{equation}
where the upper and lower signs correspond to the swimmer inside the spherical container and around the spherical obstacle, respectively.
 Therefore, the drag coefficient is given at the leading order of the asymptotic regime $X \ll R$ by 
 \begin{equation}
     M_{\rm int}=
    -\frac{9a Xl}{4 R^3}\left( 1+2\frac{X^2}{R^2}\right)
     \label{eq:Msphere1}
 \end{equation}
for the swimmer in the spherical container ($X<R$)[FIG. \ref{fig:swimmer}(a)] by neglecting the $O\left( (X/R)^4\right)$ term. With an external sphere ($X>R$)[FIG. \ref{fig:swimmer}(c)], the drag coefficient is similarly evaluated for $X \gg R$ as follows:
  \begin{equation}
  M_{\rm ext}=
  \frac{9a Rl}{4 X^3}.
  \label{eq:Msphere2}
  \end{equation}

Owing to hydrodynamic interactions, it is generally not tractable to provide an analytic expression for the function $M(X, l)$ when the swimmer is situated in the vicinity of external flat no-slip boundaries. To capture the general tendency of the dynamics, we expand the function $M(X, l)$ in the Taylor series of $X$, where $X$ represents the distance from the boundary edge, and we assume nonsingular behavior at the boundary. As the convexity of $M$ is crucial in nonzero displacements, the simplest model of $M$ should be a quadratic function with respect to $X$. Thus, we introduce a model function
 \begin{equation}
     M_{\rm quad}=\lambda^{-3}lX^2, \label{eq:Mquad}
 \end{equation}
where $\lambda$ represents the length scale of the variation of $M$.
 
In the following sections, for computational reasons we focus on the dynamics using $M_{\rm int}$ and $M_{\rm quad}$, which are both nonsingular in $X$.

  \section{\label{sec:fundmental} Reciprocal swimming in a noisy environment}

  We now proceed to examine the mechanism through which the reciprocal swimmer generates net locomotion under geometrical confinement in a noisy environment. The arguments presented in this section do not depend on the specific form of $M(X, l)$, as discussed in Sec.\ref{sec:derivation}; thus, they hold for a general reciprocal swimmer.%\yh{両方考慮して遊泳可能 }

\begin{figure}
      \centering
      \begin{overpic}
       [width=0.99\linewidth]{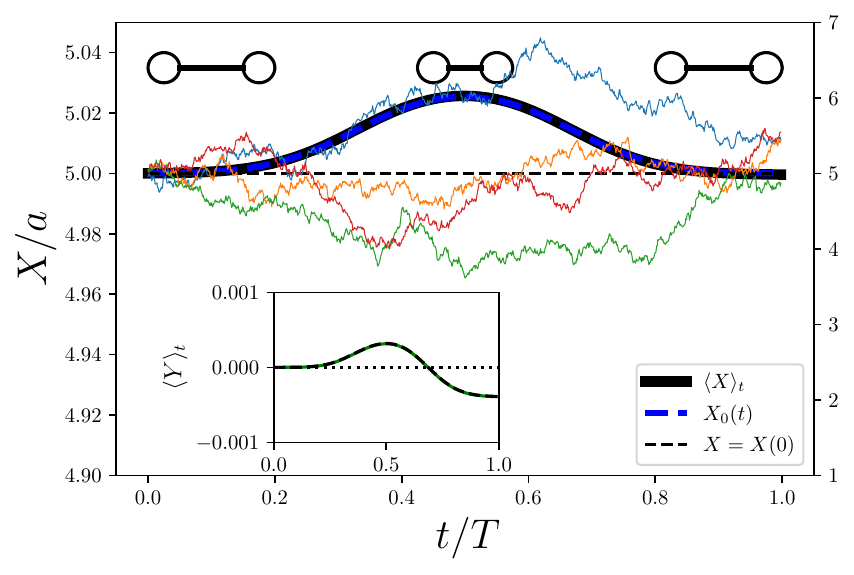}
      \put(0,62){{\large (a)}}
      \end{overpic}\\
      \begin{overpic}
     [width=0.99\linewidth]{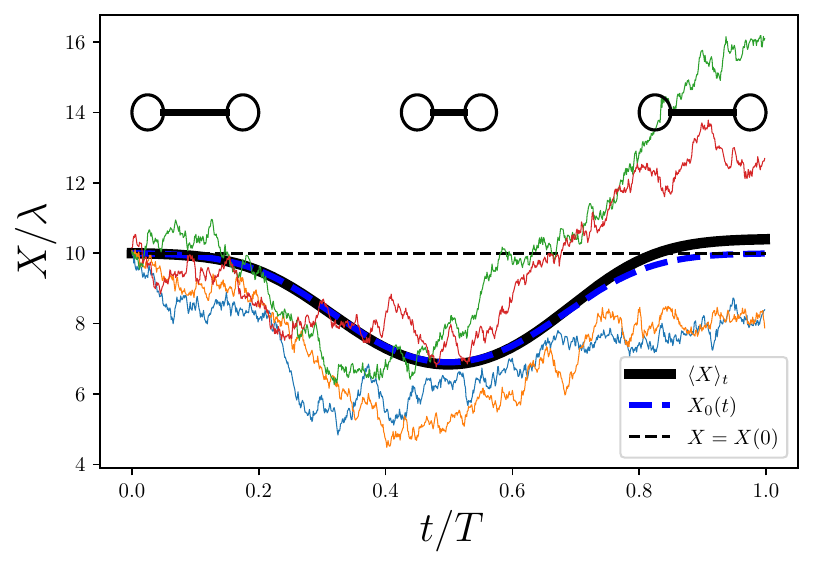}
      \put(0,62){{\large (b)}}
      \end{overpic}      
      \caption{
      Sample paths of the position of a reciprocal swimmer in a noisy environment and their statistical averages for (a) the swimmer in a spherical container $M_{\rm in}$ and (b) the model of $M_{\rm quad}=\lambda^{-3}lX^2$.
      During one reciprocal deformation, as illustrated by the schematics, the arm is shortened in the first half and elongated in the second half. 
      (a) With a symmetric reciprocal deformation cycle $l_1(t)$ ($t\in[0, 1]$), the statistical average $\langle X \rangle_t$ (shown by the thick line) deviates from the zero-noise solution $X_0$ (shown as a dashed line) in the last part of the deformation cycle, although these two almost overlap.
      [inset] Deviation from the zero-noise solution $\langle Y\rangle_t=\langle X \rangle_t -X_0(t)$ is shown by black dashed line. 
      The averaged displacement $\langle Y\rangle_{t=T}$  is clearly generated after one period of deformation at $t/T=1$. 
      The ensemble average of $10^8$ samples of the Langevin simulations is plotted by the green line, 
      although this almost overlaps with the plot of $Y(t)$.
      (b) Similar demonstrations of the deviation using a model of $M_{\rm quad}$ to indicate the behaviors with high clarity. The function of $l_1(t)$ and the parameters used in the two panels are given in the main text.
      }
      \label{fig:path_s}
    \end{figure}

In FIG.\ref{fig:path_s}, we present a numerical demonstration of the paths of a reciprocal swimmer following Eq. \eqref{eq:EOM}.
  FIG. \ref{fig:path_s} presents the individual and averaged dynamics of the swimmer inside a sphere.
  We employed the Euler--Maruyama method \cite{Maruyama1955,Kloeden1992} for the time integration over $t\in[0, T]$
  as well as the drag coefficient of Eq. \eqref{eq:Msphere1}. The shape gait is given by a symmetric form as follows: 
  \begin{equation}
      l_1(t)= l_0+\epsilon_1\exp\left[-\left(\frac{t}{T}-0.5\right)^2/s^2\right]
      \label{eq:l1},
  \end{equation}
   where parameter $s$ is fixed as $s=0.2$.

  In FIG.\ref{fig:path_s}(a), we plot the results for a swimmer in a spherical container using the expression $M_{\rm in}$\eqref{eq:Msphere1}. 
  The parameters are set as $R/a=10$, $X(0)/a=5$, $DT/a^2=1$, $l_0/a=2$, and $\epsilon_1/a=-1$. 
  The rod between the spheres is shortened and subsequently elongated to recover the initial length at time $t=T$, as indicated in the figure.
  Each realization of the stochastic dynamics is shown using differently colored thin lines. The statistical average $\langle X\rangle_t$ is computed based on the Fokker--Planck equation \eqref{eq:FP} and shown as a thick line.
  Furthermore, we plotted the zero-noise solution as a dashed line and observed that the statistical average deviates from the zero-noise solution, although the two lines appear to overlap. To show the difference, in the inset, we plotted the deviation from the zero-noise  defined as $Y(t)\equiv \langle X \rangle_t-X_0(t)$. This confirmed that the statistical average deviated from the zero-noise solution $X_0(t)$ in the last part of the deformation, while the zero-noise solution $X_0(t)$ returned to the initial position after the deformation cycle. The displacement by the reciprocal swimmer after one beat cycle is computed as $\langle X\rangle_{t=1}\approx 5.0\times 10^{-4}a$ [inset of FIG.\ref{fig:path_s}(a)]. 
  In the inset, we also plotted the ensemble average of the $10^8$ samples from the Langevin simulation of Eq. \eqref{eq:EOM}. 
  The two plots almost overlap, validating the numerical simulation of the Fokker--Planck equation with the space discretization set as $0.01a$.
  We confirmed by doubling spatial resolution that the absolute numerical error is bounded by $10^{-6}a$ and that the numerical error at time $t/T=1$ is bounded $10^{-9}a$, which is less than 0.001\% of $5\times10^{-4}a$. 

  To highlight the nonzero displacement, we also consider the example model of $M_{\rm quad}=\lambda^{-3}lX^2$, which was introduced in Sec.\ref{sec:CalcM}.
  The simulation results are presented in FIG. \ref{fig:path_s}(b) for the same stroke of $l_1(t)$, with the parameters being set as 
  $X(0)/\lambda=10$, $DT/\lambda^2=10$, $l_0/\lambda=0.5$, and $\epsilon_1/\lambda=-0.1$. 
  As is the case for a swimmer inside a spherical container, the spatial dependence of the drag coefficient produces a nonzero averaged displacement for a reciprocal swimmer in a noisy environment.

    With regard the diffusion process of a passive rigid particle under thermal noise, by dropping the first term in Eq. \eqref{eq:FP}, we obtain the simple diffusion equation. In the presence of a wall, the nonpenetrating boundary condition leads to the probabilistic flux moving away from the wall, with its averaged position evolving in time as $O(\sqrt{t})$. Notably, this diffusive displacement is unidirectional. 
    As seen in FIG.\ref{fig:path_s}, when the swimmer deforms, the statistical average of the displacement can have a nonzero value.
    As the swimmer repeats its deformation, the net displacement is expected to be proportional to time. Thus, it scales as $\langle X \rangle_t =O(t)$, which becomes much larger than the passive diffusion of $O(\sqrt{t})$ after a longer time. This physically intuitive argument is further numerically and theoretically demonstrated in the next section.

  \section{small-deviation theory\label{sec:rtheory}}
  
  To derive a formula for calculating the deviation from the zero-noise solution, we introduce an imaginary probe force $f_p$ to the system as an arbitrarily small external force instead of the environmental noise. We focus on the short-time behavior, in which the deviation from the zero-noise solution $Y= X(t)-X_0(t)$ is sufficiently small to expand the equations of motion \eqref{eq:EOM}. Thus, we obtain
    \begin{align}
      \frac{dY}{dt}=&M'(X_0(t),l(t))\frac{dl}{dt}Y \nonumber \\
      &+\frac{1}{2}M''(X_0(t),l(t))\frac{dl}{dt}Y^2+f_p(t)
      \label{eq:Yprobe} 
    \end{align}
    with $O(Y^3)$ errors, where the primes denote the derivative with respect to $X$.
    We may readily solve Eq. \eqref{eq:Yprobe} using a standard method for a linear ordinary differential equation as follows:
      \begin{align}
        Y(t)=&\int_{0}^{t}du\exp(\mathcal{L}(t,u)) \nonumber \\
        \times &\left(f_p(u)+\frac{1}{2}M''(X_0(u),l(u))\frac{dl}{du}Y(u)^2\right) \label{eq:ler1},
      \end{align}
    where the kernel $\exp\mathcal{L}$ corresponds to an impulse response. The two-time function $\mathcal{L}(t, u)$ is explicitly given as follows:
    \begin{equation}
     \mathcal{L}(t,u)=\int^{t}_{u}ds M'(X_0(s),l(s))\frac{dl}{ds}\label{eq:ler2},
    \end{equation}    
    which we now call a {\it response generator }.
    
    Applying white Gaussian noise as the probe force such that $f_p(t)=\sqrt{D}\xi(t)$ and taking the statistical average, we first obtain the variance up to the second order of $Y$, as follows:
    \begin{align}
      \langle Y\rangle_t &=\frac{1}{2}\int_{0}^{t}du\exp(\mathcal{L}(t,u))M''(X_0(u),l(u))\frac{dl}{du}\langle Y^2\rangle_{u}.\label{eq:thave}
    \end{align}
    Thereafter, we derive the expression for $ \langle Y^2\rangle_t$ by  substituting the form of the probe force $f_p(t)=\sqrt{D}\xi(t)$ into $Y^2(t)$ after squaring Eq. \eqref{eq:ler1}. Thus, the statistical average becomes
    \begin{align}
        \langle Y^2\rangle_t =&\int_{0}^{t}du\int_{0}^{t}ds\exp(\mathcal{L}(t,u)+\mathcal{L}(t,s))D\langle \xi(u)\xi(s)\rangle 
        \label{eq:Y2A}
      \end{align}    
    by neglecting the $O(Y^4)$ term and dropping other terms using $\langle \xi(u)\rangle=0$. Based upon the relation $\langle \xi(u)\xi(s)\rangle=\delta(u-s)$, we further simplify the equation to obtain
    \begin{align}
      \langle Y^2\rangle_t &=D\int_{0}^{t}du\exp(2\mathcal{L}(t,u))\label{eq:thvar}.
    \end{align}

    Equations \ref{eq:thave}) and (\ref{eq:thvar}) provide a closed form, thereby enabling us to calculate the statistical averages when the swimming gait $\dot{l}$ and its drag coefficient $M(X, l)$ are given. From Eq. (\ref{eq:thave}), the zero net displacements for the deterministic swimmer ($D=0$) and rigid swimmer without deformation ($dl/dt=0$) can be easily reproduced.

    Let us further assume that the rod length $l(t)$ is a periodic function with a period $T$. Thus, the integrands in 
    Eqs.(\ref{eq:thave}) and (\ref{eq:thvar}) as well as the response generator $\mathcal{L}$ are also periodic functions with the same period. Therefore, we can introduce the net swimming velocity and net diffusion coefficient using the time-average of $\langle Y \rangle_t$ and $\langle Y^2 \rangle_t$ over the deformation cycle. 

    Here, we define the net swimming velocity as
    \begin{align}
      U_{\rm net}=\frac{\langle Y\rangle_{nT}}{nT} \label{eq:vele},
    \end{align}
    with a positive integer $n$. This velocity is equivalent to the net swimming velocity because the zero-noise solution vanishes as $X_0=0$ at time $t=nT$.
    Similarly, we define the net diffusion coefficient as
    \begin{align}
      D_{\rm net}=\frac{\langle Y^2\rangle_{nT}}{nT} \label{eq:vare},
    \end{align}
    which is explicitly obtained from Eq. \eqref{eq:thvar} as
    \begin{align}      
      D_{\rm net}&=\frac{D}{T}\int_0^Tdt\exp(-2\mathcal{L}(t,0)).\label{eq:Dnet}
    \end{align}

    In the subsequent sections, we employ Eqs. (\ref{eq:thave}) and (\ref{eq:thvar}) to examine the statistical properties of the swimmer.

  \section{Net swimming velocity and diffusion of the swimmer}
  \label{sec:VelocAndDiff}

  \subsection{Net swimming velocity\label{sec:velocity}}

  Next, we analyze the ensemble-averaged swimming velocity and variance using the formula derived in the previous section.

  \begin{figure}
    \centering
    \begin{overpic}
    [width=0.79\linewidth]{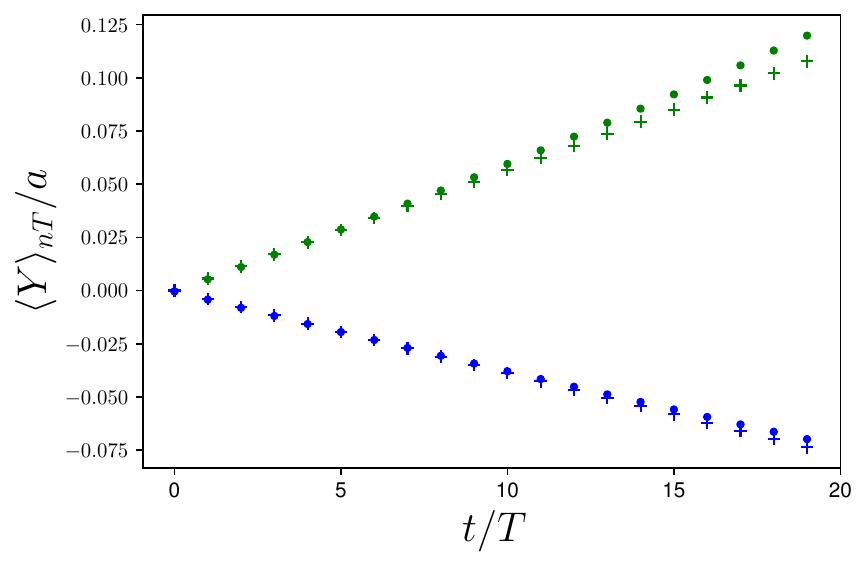}
    \put(-2,62){{(a)}}
    \end{overpic}
    \begin{overpic}[width=0.79\linewidth]{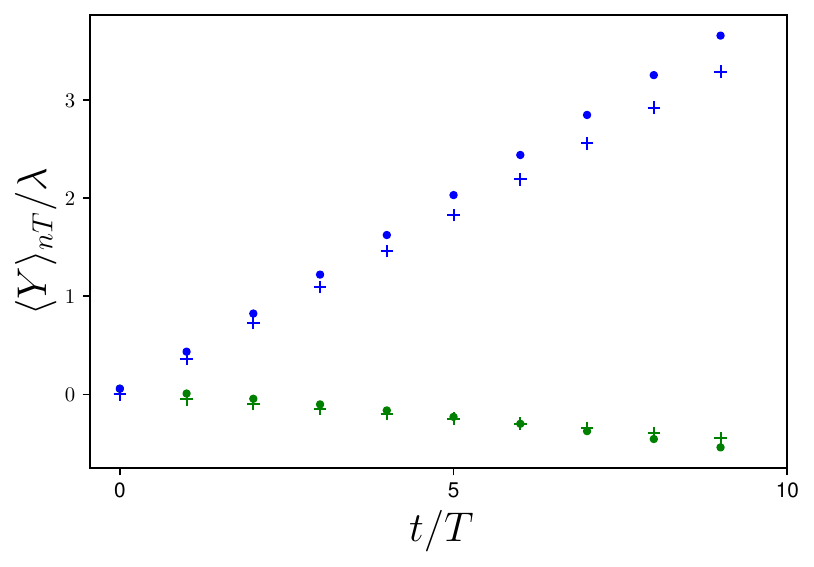}
    \put(-2,62){{(b)}}
    \end{overpic}
    \caption{Net displacement at every deformation period ($t/T=0,1,\cdots$) for (a) a reciprocal swimmer in a spherical container with drag coefficient $M_{\rm int}$ given by Eq. \eqref{eq:Msphere1} and (b) an example model with $M_{\rm quad}$ given by Eq. \eqref{eq:Mquad}. 
    We employed the shape gait of $l_1(t)$ [Eq. \eqref{eq:l1}] in both cases.
    (a) The circles ($\circ$) indicate the numerical results, and the plus symbols ($+$) indicate the theoretical predictions obtained using the small-deviation theory [Eq. \eqref{eq:thave}]. The results with $\epsilon_1>0$ and $\epsilon_1<0$ are represented by the green (increasing) and blue (decreasing) points, respectively. The numerical simulation and prediction based on the small-deviation theory are in excellent agreement. (b) Similar plots for the net displacement with $M_{\rm quad}$.
    \label{fig:XSpoi}}
  \end{figure}

  We now analyze the net velocity for particular swimmer models from the formulas [Eqs. \eqref{eq:thave} and \eqref{eq:thvar}] for comparison with the solution of the Fokker--Planck equation. As in FIG.\ref{fig:path_s}, we employ the symmetric deformation $l(t)=l_1(t)$ for $t\in[0, T]$ and recursively repeat this deformation at a later time using $l(t+T)=l(t)$. Other parameters of the shape gait were the same as those shown in FIG. \ref{fig:path_s}, except the sign of $\epsilon_1$ and the value of the diffusion constant, which was set to $DT/a^2=10$.
  
  In FIG.\ref{fig:XSpoi}, we plot the averaged deviation from the zero-noise solution or, equivalently, the net displacement $\langle Y\rangle$ at every period of deformation. FIG. \ref{fig:XSpoi}(a) presents the analysis results for $M_{\rm int}$.
  The green and blue plots show the results with $\epsilon_1/a=1$ and $\epsilon_1/a=-1$, respectively. 
  The results from the formulas [Eqs. \eqref{eq:thave} and \eqref{eq:thvar}] plotted using the plus symbol ($+$) are in excellent agreement with the solution to the Fokker--Planck equation plotted by circles ($\circ)$. The displacement is linear in time both in the theoretical prediction and in the direct simulation of the Fokker--Planck equation.

  Here, we again emphasize that the linear increase of the averaged displacement dominates over the passive diffusion of $O(\sqrt{t})$ at large times. Further, upon changing the shape gait by $\epsilon_1 \rightarrow -\epsilon_1$, the swimmer can reverse its direction, being distinct from the unidirectional diffusion of passive particles in the presence of an external boundary.

  To investigate the limit of our theory based on the small-deviation theory, we again examine the model of $M_{\rm quad}$, and the same parameter set is used as shown in FIG.\ref{fig:path_s}(b). 
  As in the study using $M_{\rm in}$, we found a linear increase in the net displacement in time, a reversal of the moving direction, and a flip of the shape gait.
  The deviation from the zero-noise solution is relatively large in this case, and the assumption of a small deviation is expected to be violated at an early time. In FIG.\ref{fig:XSpoi}(b), we plotted the net displacement at every deformation period. The statistical average obtained through direct numerical simulation deviates from that obtained through theoretical prediction, which assumed a small deviation and neglected $O(Y^4)$ terms.

  \subsection{Net swimming velocity and shape gait\label{sec:deform}}
  Next, we examine the effects of the shape gait on locomotion by focusing on the short-time regime, where the small-deviation theory is valid. 
  Notably, the function $M$ was proportional to $l$ in all four examples of $M_{\textrm{flat}}$, $M_{\textrm{ext}}$, $M_{\textrm{int}}$, and $M_{\textrm{quad}}$; therefore, we decomposed $M$ into the form $M(X(t),l(t))=N(X(t))l(t)$.
  Further, in the short-time regime under consideration, the function $N$ is represented by the value at the initial time and we can approximate its second spatial derivative as $N''(X(t))=N''(X(0))+O(N'''(X(t)-X(0)))$, leading to the following relation:
     \begin{equation}
       M''\frac{dl}{dt}=\frac{\kappa}{2} \frac{dl^2}{dt}+O(N'''(X(t)-X(0)))
       \label{eq:kappa},
   \end{equation}
    where $\kappa=N''(0)$ is a constant determined only by the initial position.
   Subsequently, we plugged these equations into Eqs. \eqref{eq:thave} and \eqref{eq:thvar}. The integrands are no longer dependent on time, so we obtain the following:
   \begin{align}
  \langle Y\rangle_t =\frac{Dt\kappa}{2}\left( (l(t)^2-\frac{1}{t}\int_0^t dt' l(t')^2\right). \label{eq:app}
\end{align}
   Next, we introduce a $T$-periodic function $\eta (t)$ such that $l(t)=l_0+\eta (t)$ and $\eta (T)=\eta(0)=0$. By substituting this form into Eq. \eqref{eq:app},
   the net velocity at $t=nT$ is given by
    \begin{align}
      U_{\rm net}
     =-\frac{D\kappa}{2}\left( \frac{2 l_0}{T}\int_0^T du \,\textrm{Sym}[\eta(u)] +\frac{1}{T}\int_0^T du \eta^2(u) \right) \label{eq:app3},
    \end{align}
    where the time-reversal part of the deformation is defined by $\textrm{Sym}[\eta(t)]\equiv (\eta(t)+\eta(T-t))/2$.
    From Eq. \eqref{eq:app3}, at the first order of $\eta$, only the time-reversal term contributes to the net velocity. 
    Similarly, we can define the skew-symmetric part of the deformation as $\textrm{Skew}[\eta]\equiv (\eta(t)-\eta(T-t))/2$, and the skew-symmetric part of the deformation rate is computed as
    \begin{equation}
    \textrm{Skew}\left[\frac{dl}{dt}(t)\right]
    = \frac{1}{2}\left(\frac{dl}{dt}(t)-\frac{dl}{dt}(T-t)\right)=\frac{d}{dt}\textrm{Sym}[\eta(t)]
    \label{eq:UnetEq}.
    \end{equation}
    
  To examine the effects of the shape gait and its symmetry, we consider two examples of the deformation function with different symmetries.
    The first example is the shape gait given by $l(t)=l_1(t)$, which was already introduced as Eq. \eqref{eq:l1} in Sec.\ref{sec:setting}, and it exhibits time-reversal symmetry as  $l_1(t)=l_1(T-t)$.
    As the second example, we consider a skew-symmetric oscillatory deformation given by 
    \begin{equation}
    l_2(t)= l_0+\epsilon_2\sin(\omega t)  
    \label{eq:l2},
    \end{equation}
    where $\omega=2\pi/T$ is the angular frequency. This function satisfies the relation $l_2(t)-l_0=-(l_2(T-t)-l_0)$.
  
  In these two cases, by substituting the shape gait functions into Eq. \eqref{eq:app3}, 
  we obtain exact solutions that estimate the net displacement during one period of deformation. 
  For the symmetric deformation $l_1(t)$, the net velocity is expressed as
  \begin{equation}
  U^{(1)}_{\textrm{net}}=-D\kappa\sqrt{\pi} \,s\left(2l_0\epsilon_1+\frac{1}{\sqrt{2}}\epsilon_1^2\right)
      \label{eq:Unet1}.
  \end{equation}
  Conversely, the expression for the skew-symmetric deformation $l_2(t)$ possesses a qualitatively different form, as follows:
  \begin{equation}
      U^{(2)}_{\textrm{net}}=\frac{1}{4}D\kappa \epsilon_2^2
      \label{eq:Unet2},
  \end{equation}   
   which is independent of $\omega$. 
   The symmetric deformation yields contributions of the order of $\epsilon_1$, whereas the skew-symmetric deformation only generates net velocities of the order of $O(\epsilon_2^2)$.

   From Eq. \eqref{eq:Unet1}, we again confirm that the deformation or the nonzero $\epsilon_1$ is necessary for generating the net displacement, which is proportional to time. Furthermore, if $|\epsilon_1/l_0|<2\sqrt{2}$, by changing the shape gait with $\epsilon_1\rightarrow-\epsilon_1$, we can invert the swimming direction.
  
\begin{figure}
    \centering
    \includegraphics[width=0.99\linewidth]{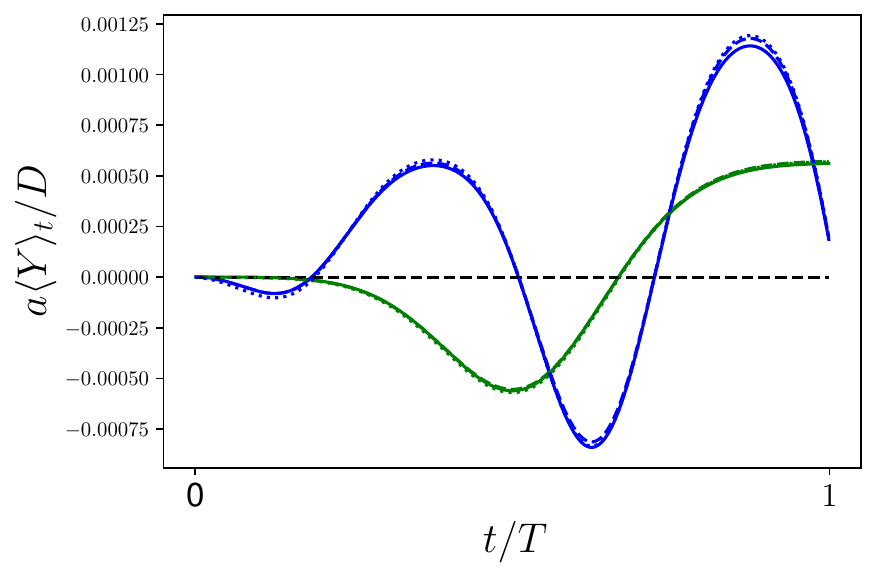}
    \caption{Net displacement from the zero-noise solution for a swimmer in a spherical container with two shape gaits, $l(t)=l_1(t)$ (blue) and $l(t)=l_2(t)$ (green).
    The solid lines indicate the theoretical prediction given by Eq. \eqref{eq:app}, and the dotted lines indicate $a\langle Y \rangle_t/D $, which is the numerical solution to the Fokker--Planck equation.
    The ensemble average of stochastic simulation of the Langevin equation is plotted by the dashed line for each shape gait, although the plots almost overlap with the Fokker--Planck solutions.
    The parameters are set as $(X_0, l_0, \epsilon_1, \epsilon_2)=(5,2,1,1)a$ and $DT/a^2=10$.
    }
    \label{fig:app}
    \end{figure}

    In FIG. \ref{fig:app}, we plot the numerical result of the net displacement via the Fokker--Planck and Langevin equations,
    along with the theoretical results for $M_{\rm int}$.
    We used the parameters $(X_0,l_0,\epsilon_1,\epsilon_2)=(5,2,1,1)a$ with the diffusion constant $DT/a^2=10$ to integrate over time $t \in [0,T]$.
    For both shape gaits, the analytical solutions are in excellent agreement with the numerical solutions, which, in turn validates the approximations for deriving Eq. \eqref{eq:app}.

  \subsection{Net diffusion and shape gait\label{sec:fluct}}

 We further investigate the relations between the shape gait and the emerging net velocity and diffusion by analyzing the impulse response function $\exp(\mathcal{L}(t_1, t_2))$. From the definition presented in Eq. \eqref{eq:ler1}, this function characterizes the effect of noise at time $t_2$ on the locomotion at time $t_1$ and is therefore interpreted as sensitivity to environmental noise. Furthermore, as expressed in Eq. \eqref{eq:thvar}, the impulse function affects $\langle Y^2\rangle$.%; hence, the net diffusion is constant .

\begin{figure}
    \centering
    \includegraphics[width=0.99\linewidth]{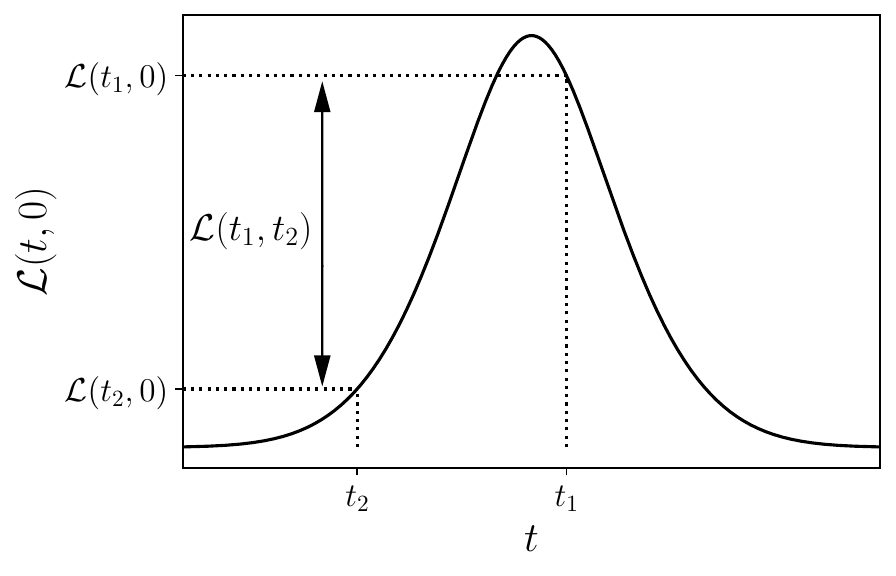}
    \caption{Illustration of the value of the response generator $\mathcal{L}(t_1,t_2)$ using a graph of $\mathcal{L}(t,0)$. 
    }
    \label{fig:L}
  \end{figure}

    To visually understand this effect, we first decompose the response generator into $\mathcal{L}(t_1,t_2)=\mathcal{L}(t_1,0)-
    \mathcal{L}(t_2,0)$. A schematic of this function $\mathcal{L}(t,0)$ is shown in FIG.\ref{fig:L}. Therefore, the response generator $\mathcal{L}(t_1,t_2)$ is regarded as the difference between the two points on the graph of $\mathcal{L}(t,0)$.

    To gain an intuitive understanding, we consider a polynomial function $M(X,l)$, where $M(X,l)=\lambda^{-(m+1)}lX^m$ and $m$ is an integer $m\neq 1$ that includes $M_{\rm flat}, M_{\rm ext}$, and $M_{\rm quad}$ as $m=-2, -3,$ and $2$, respectively. 
    
    Notably, the expressions derived in Sec. \ref{sec:derivation} are written as a sum of polynomial functions. With this simplified form, we can exactly solve the response generator $\mathcal{L}(t,0)$. Expectedly, substituting the form of $M$ and integrating \eqref{eq:EOMdet} lead to the expression of the zero-noise solution, as follows:
    \begin{equation}
     X_0(t)=\left[ \frac{(1-m)}{2}\lambda^{-(m+1)}l^2(t)+c\right]^{1/(1-m)}
     \label{eq:ExactSlnX},
    \end{equation}
    where $c$ is the constant of the integral given by $c=X(0)^{(1-m)}-(1-m)Kl^2(0)/2$. Thereafter, using this expression to integrate Eq. \eqref{eq:ler2}, we obtain the response generator in the following form:
    \begin{equation}
        \mathcal{L}(t,0)=m\log\left(
        \frac{X_0(t)}{X_0(0)}\right)
        \label{eq:ExactSlnL}.
    \end{equation}

  Notably, when $m>1$, the shape gait $l(t)$ cannot be arbitrarily taken. % bounded by the constraint that the zero-noise solution is bounded. 
    As the first term in the bracket of Eq. \eqref{eq:ExactSlnX} is negative, the second term $c$ needs to be positive, leading to a diverging $X_0$ parameter at some finite values of $l^2(t)$.

\begin{figure}[!t]
    \centering
    \begin{overpic}[width=0.79\linewidth]{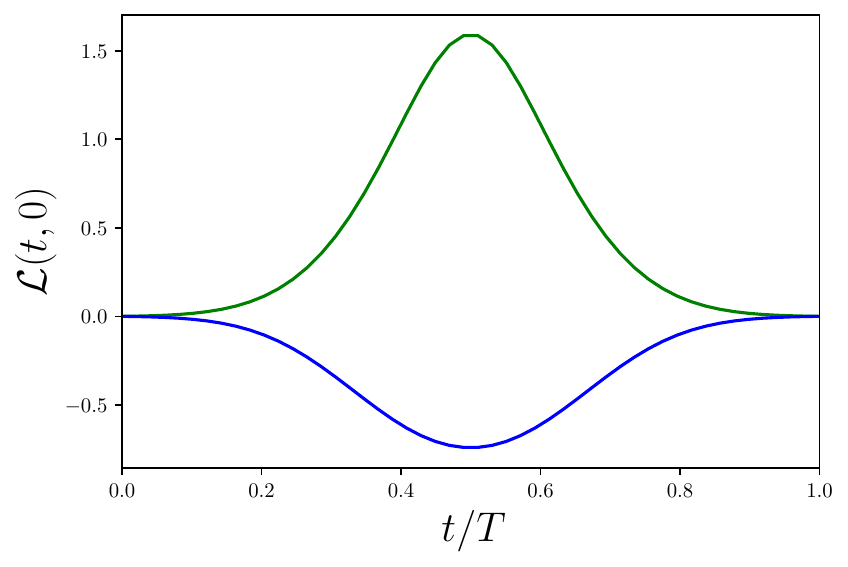}
    \put(-2,62){{(a)}}
    \end{overpic}
    \begin{overpic}[width=0.79\linewidth]{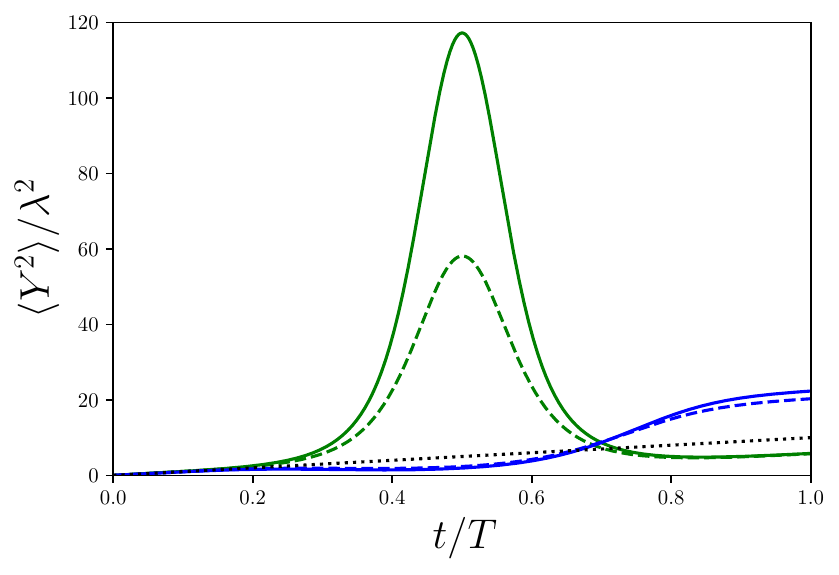}
    \put(-2,62){{(b)}}
    \end{overpic}    
    \caption{
    (a) Plots of $\mathcal{L}(t,0)$ obtained from the exact solution \eqref{eq:ExactSlnL} with $M_{\rm quad}$. The shape gait is given by $l_1(t)$, and the parameters are the same as shown in FIG.\ref{fig:XSpoi}(b). The green and blue plots correspond to cases with $\epsilon_1=0.1\lambda$ and $\epsilon_1=-0.1\lambda$, respectively.
    (b) Variance $\langle X^2 \rangle_t$ for the same swimming gaits. 
    The solid lines indicate the results from direct computation, and 
    the dashed lines show the theoretical predictions obtained using Eq. \eqref{eq:thvar}.
    We compare these lines with the normal diffusion $Dt/\lambda^2$ plotted by the black-dotted line. 
    For both signs of $\epsilon_1$, the shape change can contribute to enhancing and suppressing net diffusion. 
    \label{fig:XS2}}
  \end{figure}

    The exact solution is plotted in FIG.\ref{fig:XS2}(a) for $m=2$ or $M_{\rm quad}$, with the initial position given as $X_0(0)=10\lambda$ and a shape gait of $l_1(t)$ in Eq. \eqref{eq:l1}. We employ the same parameter set as in FIGs.\ref{fig:path_s}(b) and \ref{fig:XSpoi}(b), where $l_0=0.5\lambda$ and $\epsilon_1=\pm 0.1\lambda$.

    When $m=2$, as in $M_{\rm quad}$, the exact solutions Eqs. \eqref{eq:ExactSlnX} and \eqref{eq:ExactSlnL} are simply written as 
    \begin{equation}
        \mathcal{L}(t,0)=-2\log\left( 1-\frac{X_0(0)}{2\lambda^3}\left(l^2(t)-l^2(0)\right)\right)
        \label{eq:ExtSlnN2}.
    \end{equation}
    Thus, the response generator becomes positive when the swimmer is elongated from the initial configuration ($\epsilon_1>0$) and negative when the swimmer is shortened from the initial configuration  ($\epsilon_1<0$). After one period of deformation at $t=T$, the response generator satisfies $\mathcal{L}(T,0)=\mathcal{L}(0,0)=0$.

    To examine the impact of the swimmer's shape gait on the variance $\langle Y^2 \rangle_t$, we numerically evaluated this function  by a direct numerical simulation of the Fokker--Planck equation and the small-deviation theory. The results are presented in FIG.\ref{fig:XS2}(b), with the normal diffusion $Dt/\lambda^2$ plotted as the black-dotted line. As shown in the figure,
    the numerical results and theoretical predictions are in good agreement. In particular, the shape deformation contributes to enhancing or suppressing the net diffusion depending on the sign of $\mathcal{L}(t_1, t_2)$, as expressed in Eq. \eqref{eq:thvar}. The response generator does not change its sign for $t<0.5 T$ because the function $\mathcal{L}(t, 0)$ is monotonic. 
    The value of the variance $\langle Y^2 \rangle_t$ therefore deviates from the normal diffusion with enhancement ($\epsilon_1>0$) or suppression ($\epsilon_1<0$) [FIG.\ref{fig:XS2}(b)].
    However, later during the deformation, the sign and impact of the impulse response on the variance $\langle Y^2 \rangle_t$ change and subsequently reverse (i.e., suppression when $\epsilon_1>0$ or enhancement when $\epsilon_1<0$), as evidenced in the figure.

    The fluctuating diffusion during one beat cycle holds for a general swimmer with an arbitrary $M(l, t)$, according to the small-deviation theory. Indeed, by integrating Eq. \eqref{eq:ler2} from $l(t=0)$ to $l(t=T)$ after a change of variable, we have $\mathcal{L}(T,0)=\mathcal{L}(0,0)=0$. Thus, the function $\mathcal{L}(t,0)$ exhibits at least one extremum in $t\in(0, T)$ if the swimmer's shape is deformed. Therefore, for a general deforming object under a noisy environment, we conclude that there exist time periods where the net diffusion exceeds the normal diffusion $D$ and where the net diffusion is suppressed more than the normal diffusion.

  \section{Concluding remarks\label{sec:conc}}
  In this study, we theoretically investigated the dynamics of a reciprocal microswimmer in a noisy environment. Focusing on a microswimmer moving in one direction, we showed that the statistical average of its displacement $\langle X \rangle_t$ can produce net locomotion if the swimmer's dynamics are confined by external boundaries. As the scallop theorem states, such a reciprocal swimmer returns to its original position after one beat cycle in the absence of environmental noise. Its dynamics were introduced as the zero-noise solution and denoted as $X_0(t)$.
  The mathematical structure behind the net displacement lies in the convexity of $M(X,l)$. Hence, the scallop theorem is expected to be violated in the statistical sense in more general situations than our assumption of one-dimensional dynamics. The small-deviation theory for multiple dimensions, however, requires a noncommutative integral for an impulse response, which yields theoretical challenges. Therefore, the details of these extensions will be reported elsewhere in the future.
  
  For a quantitative analysis of net locomotion in a statistical sense based on an impulse response function, we established a theory to analyze the deviation from the zero-noise solution, given by $\langle Y\rangle_t=\langle X \rangle_t-X_0(t)$. 
  Focusing on a two-sphere model swimmer located in a spherical container, we numerically and theoretically demonstrated that the reciprocal swimmer can result in net displacement in a statistical sense; the displacement linearly increases as time progresses. This dominates the unidirectional diffusion of passive particles, with its net displacement scaling as $O(\sqrt{t})$ after a long time. 
  Reciprocal swimming should be enhanced near a wall, which can be estimated as $M_{\rm{ quad}}$.
  This provides a net velocity of $U_{\textrm{net}}\approx 0.47\mu$m$/$s, where we use the physical units of $\lambda=a=1\mu $m, $\Theta=300$K, and $T=1$s. 
  We may also estimate the characteristic timescale $t^{\ast}$ at which the active locomotion is of the same order in magnitude as the passive diffusion. With the same parameter sets, from $U_{\rm net}t^{\ast}=\sqrt{Dt^{\ast}}$, we obtain $t^{\ast}\approx 2.2$s. 
   Further, we demonstrate that by controlling its shape gait, the swimmer can migrate in both directions, which is remarkably different from the unidirectional diffusive motion of passive particles.

  Although our theory is based on a small deviation over a short timescale, it is sufficient to analyze the key mathematical structures of the shape gait that produce net statistical locomotion. 
  Based on this theory, we found that a time-reversal deformation produces the net swimming velocity at the leading order, whereas the skew-symmetric part of the deformation only contributes to the second order of the deformation amplitude. Further, we found that the shape gait affects the net diffusion constant through the behavior of the impulse response function and demonstrated that net diffusion is both enhanced and suppressed during one beat cycle.

The perturbation method used in this study is similar to that used for nonlinear phenomena with very high dimensions, such as high-Reynolds-number turbulence.
The high-order moments of the probability distribution function are generally not negligible in these systems. Thus, further theoretical treatments are required to include these moments, such as direct interaction approximation in turbulent flows \cite{Kraichnan1959,kida1997lagrangian, Goto2002a}
and the renormalization group theory in quantum field theory.
In contrast, the perturbation analysis around a given nonlinear solution, as used in this study, does not require high-order moments, for which our small-deviation theory was matched with the full dynamics
\cite{pontryagin2014ordinary}.

Another remark on the small-deviation theory is that it is limited to short-time behavior (i.e., when the deviation from the zero-noise solution is sufficiently small). Long-time asymptotic behavior is also important because it can be used to predict the accumulation or depletion of swimmers around the external boundaries,
%In these situations , the deformation-driven net locomotion of a reciprocal microswimmer is observable and significant in physical and %biological problems, 
such as the surface accumulation of microswimmers \cite{ishimoto2019bacterial} as well as the fluctuation of polymers and enzymes inside biological and artificial cell membranes \cite{gnesotto2018broken}.
Numerical works with such a singular function, however, require careful treatment to satisfy the no-penetration condition at the boundary. Further, precise cell--wall hydrodynamic interactions frequently incur considerable numerical cost because of the fine numerical meshes \cite{ishikawa2006hydrodynamic}.

In conclusion, we theoretically and numerically studied the fundamental features of the statistical properties of a microscopic object with reciprocal deformation in a fluctuating and geometrically confined environment. We focused on the short-time, small-deviation situation to elucidate the nontrivial, deformation-driven net locomotion of a reciprocal swimmer.
Our findings on the relation between the symmetry of the shape gait  and the net locomotion and the theoretical formulation obtained based on the impulse response provide a fundamental basis for environment-coupled statistical locomotion. Thus, this study will be beneficial for understanding biophysical dynamics in fluctuating environments, designing artificial microrobots, and conducting laboratory experiments.

\section*{Acknowledgments}
K.I. acknowledges the Japan Society for the Promotion of Science (JSPS) KAKENHI for Transformative Research Areas A (Grant No. 21H05309) and the Japan Science and Technology Agency (JST), FOREST (Grant No. JPMJFR212N). Y.H. and K.I. were supported in part by the Research Institute for Mathematical Sciences, an International Joint Usage/Research Center located at Kyoto University.\\

\appendix
\section{Derivation of Eq. \eqref{eq:EOM}}
\label{app:derivEoM}
    In this appendix, we provide an error estimate to derive the equation of motion \eqref{eq:EOM}.
    The balance among the drag force, propulsion, and external noise force,
    $F_{\textrm{d}}+F_{\textrm{p}}+F_{\textrm{ext}}=0$, should be formulated by a stochastic differential equation of the Stratonovich type, as follows:
    \begin{equation}
    C(X, t)dX=A(X, t)dt+B(X,t)dW
        \label{eq:App11},
    \end{equation}
    where $C(x, t)=2\gamma$ and $A=F_{\textrm{p}}$ from Eqs. \eqref{eq:Fd2}--\eqref{eq:Ft2}; $dW$ indicates a Wiener process. 
    Thus, the corresponding equation motion in the sense of It\^o is given by
    \begin{equation}
    \left( C+\frac{C'}{2}dX\right) dX=Adt+\left( B+\frac{B'}{2}dX\right) dW
        \label{eq:App12},
    \end{equation}
    where the prime denotes a derivative with respect to $X$. With It\^o's rules, Eq. \eqref{eq:App12} is rewritten in the order of $dt$ as
    \begin{equation}
        C(X, t)dX=\left( A+ \frac{BB'}{2}-\frac{C'B^2}{2C^2}\right)dt+BdW
        \label{eq:App13}.
    \end{equation}
    By introducing the diffusion $D$ as $D=B^2/C^2$, we arrive at the equation of motion in the following form:
    \begin{equation}
        \frac{dX}{dt}=\frac{F_{\textrm{p}}}{2\gamma}+\left( \frac{1}{2\gamma}\frac{d}{dX}(\gamma^2 D)-\sqrt{D}\frac{d\gamma}{dX}\right)+\sqrt{D}\xi,
        \label{eq:App14}
    \end{equation}
    where $\xi$ is the zero-mean Gaussian noise, which satisfies $\langle \xi(t) \rangle=0$ and $\langle \xi(t_1)\xi(t_2)\rangle \delta(t_1-t_2)$, as in the main text.
    The respective terms on the right-hand side of Eq. \eqref{eq:App14} represent the deterministic swimming, effective drift, and noise-induced velocities. 
    
    Let us rewrite Eq. \eqref{eq:App14} as $dX/dt=V_{\textrm{s}}+V_{\textrm{p}}+V_{\textrm{n}}$ and evaluate the magnitude of each velocity. 
    We first consider the ratio of the drift velocity to swimming velocity. 
    With the expressions of $V_{\textrm{d}}=-\frac{D\gamma'}{2\gamma}$ and $F_{\textrm{ p}}=-\gamma^2(\tilde{G}_{11}-\tilde{G}_{22})(dl/dt)$ [Eq. \eqref{eq:Ft2}], we estimate the ratio by neglecting prefactors as follows:
    \begin{equation}
    \frac{V_{\textrm{d}}}{V_{\textrm{s}}}\sim
    \frac{V_{\textrm{d}}}{\gamma^2\tilde{G}(dl/dt)}\sim \frac{\tilde{G}'}{\tilde{G}}\frac{D}{(dl/dt)}\sim \frac{D}{L_W (a/T)}
    \label{eq:App15},
    \end{equation}
    where $\tilde{G}$ represents the characteristic size of $\tilde{G}_{11}-\tilde{G}_{22}$. $T$ is the time of the deformation, and $L_W$ is the distance between the swimmer and wall boundary of the geometrical confinement, as introduced in the main text. Here, we define the characteristic timescale of the (passive) diffusion as $\tau_D=a^2/D$. Therefore, the ratio of the two terms is evaluated as
    \begin{equation}
    \frac{V_{\textrm{d}}}{V_{\textrm{s}}}\sim \frac{T}{\tau_D}\frac{a}{L_W}\ll 1
        \label{eq:App16},
    \end{equation}
    which is asymptotically small.
    
    The ratio of the swimmer and the noise term, in contrast, is evaluated using the displacement during the time interval $\Delta t\sim T$ and spatial interval $\Delta l$ as follows:
    \begin{equation}
    \frac{V_{\textrm{n}}}{V_{\textrm{s}}}\sim \frac{\sqrt{D \Delta t}}{\gamma \tilde{G}\Delta l}\sim\frac{\sqrt{D}}{a}\frac{\sqrt{\Delta t}}{\gamma \tilde{G}}
    \sim\frac{\sqrt{T}}{\sqrt{\tau_D}}\frac{1}{M}
    \label{eq:App17}.
    \end{equation}
    Although the magnitude of $M$ depends on the geometrical confinement, for a swimmer in a spherical container, it can be estimated from Eq. \eqref{eq:Msphere1} as $M\sim a^2 X /L_W^3\ll 1$, indicating that the noise term is dominant.

    Notably, Sancho et al. \cite{Sancho1982} also suggested using the Stratonovich product for the drag force, as in the left-hand side of Eq. \eqref{eq:App11}. However, based on physical reasoning, they suggested using the It\^o product in the stochastic part of Eq. \eqref{eq:App11}. Nonetheless, this difference in interpretation only affects the prefactor of the drift velocity; the estimate argued in this appendix remains unchanged.

    In summary, in our asymptotic regime, the drift terms are negligible; thus, we arrive at our equation of motion \eqref{eq:EOM}.

  \bibliographystyle{apsrev4-2}
  \bibliography{bib}

\begin{thebibliography}{65}%
\makeatletter
\providecommand \@ifxundefined [1]{%
 \@ifx{#1\undefined}
}%
\providecommand \@ifnum [1]{%
 \ifnum #1\expandafter \@firstoftwo
 \else \expandafter \@secondoftwo
 \fi
}%
\providecommand \@ifx [1]{%
 \ifx #1\expandafter \@firstoftwo
 \else \expandafter \@secondoftwo
 \fi
}%
\providecommand \natexlab [1]{#1}%
\providecommand \enquote  [1]{``#1''}%
\providecommand \bibnamefont  [1]{#1}%
\providecommand \bibfnamefont [1]{#1}%
\providecommand \citenamefont [1]{#1}%
\providecommand \href@noop [0]{\@secondoftwo}%
\providecommand \href [0]{\begingroup \@sanitize@url \@href}%
\providecommand \@href[1]{\@@startlink{#1}\@@href}%
\providecommand \@@href[1]{\endgroup#1\@@endlink}%
\providecommand \@sanitize@url [0]{\catcode `\\12\catcode `\$12\catcode
  `\&12\catcode `\#12\catcode `\^12\catcode `\_12\catcode `\%12\relax}%
\providecommand \@@startlink[1]{}%
\providecommand \@@endlink[0]{}%
\providecommand \url  [0]{\begingroup\@sanitize@url \@url }%
\providecommand \@url [1]{\endgroup\@href {#1}{\urlprefix }}%
\providecommand \urlprefix  [0]{URL }%
\providecommand \Eprint [0]{\href }%
\providecommand \doibase [0]{https://doi.org/}%
\providecommand \selectlanguage [0]{\@gobble}%
\providecommand \bibinfo  [0]{\@secondoftwo}%
\providecommand \bibfield  [0]{\@secondoftwo}%
\providecommand \translation [1]{[#1]}%
\providecommand \BibitemOpen [0]{}%
\providecommand \bibitemStop [0]{}%
\providecommand \bibitemNoStop [0]{.\EOS\space}%
\providecommand \EOS [0]{\spacefactor3000\relax}%
\providecommand \BibitemShut  [1]{\csname bibitem#1\endcsname}%
\let\auto@bib@innerbib\@empty
%</preamble>
\bibitem [{\citenamefont {Friedrich}\ and\ \citenamefont
  {J{\"u}licher}(2007)}]{friedrich2007chemotaxis}%
  \BibitemOpen
  \bibfield  {author} {\bibinfo {author} {\bibfnamefont {B.~M.}\ \bibnamefont
  {Friedrich}}\ and\ \bibinfo {author} {\bibfnamefont {F.}~\bibnamefont
  {J{\"u}licher}},\ }\href@noop {} {\bibfield  {journal} {\bibinfo  {journal}
  {Proceedings of the National Academy of Sciences}\ }\textbf {\bibinfo
  {volume} {104}},\ \bibinfo {pages} {13256} (\bibinfo {year}
  {2007})}\BibitemShut {NoStop}%
\bibitem [{\citenamefont {Shiba}\ \emph {et~al.}(2008)\citenamefont {Shiba},
  \citenamefont {Baba}, \citenamefont {Inoue},\ and\ \citenamefont
  {Yoshida}}]{shiba2008ca2+}%
  \BibitemOpen
  \bibfield  {author} {\bibinfo {author} {\bibfnamefont {K.}~\bibnamefont
  {Shiba}}, \bibinfo {author} {\bibfnamefont {S.~A.}\ \bibnamefont {Baba}},
  \bibinfo {author} {\bibfnamefont {T.}~\bibnamefont {Inoue}},\ and\ \bibinfo
  {author} {\bibfnamefont {M.}~\bibnamefont {Yoshida}},\ }\href@noop {}
  {\bibfield  {journal} {\bibinfo  {journal} {Proceedings of the National
  Academy of Sciences}\ }\textbf {\bibinfo {volume} {105}},\ \bibinfo {pages}
  {19312} (\bibinfo {year} {2008})}\BibitemShut {NoStop}%
\bibitem [{\citenamefont {Desai}\ \emph {et~al.}(2019)\citenamefont {Desai},
  \citenamefont {Shaik},\ and\ \citenamefont
  {Ardekani}}]{desai2019hydrodynamic}%
  \BibitemOpen
  \bibfield  {author} {\bibinfo {author} {\bibfnamefont {N.}~\bibnamefont
  {Desai}}, \bibinfo {author} {\bibfnamefont {V.~A.}\ \bibnamefont {Shaik}},\
  and\ \bibinfo {author} {\bibfnamefont {A.~M.}\ \bibnamefont {Ardekani}},\
  }\href@noop {} {\bibfield  {journal} {\bibinfo  {journal} {Frontiers in
  microbiology}\ }\textbf {\bibinfo {volume} {10}},\ \bibinfo {pages} {289}
  (\bibinfo {year} {2019})}\BibitemShut {NoStop}%
\bibitem [{\citenamefont {Hillesdon}\ and\ \citenamefont
  {Pedley}(1996)}]{hillesdon1996bioconvection}%
  \BibitemOpen
  \bibfield  {author} {\bibinfo {author} {\bibfnamefont {A.}~\bibnamefont
  {Hillesdon}}\ and\ \bibinfo {author} {\bibfnamefont {T.}~\bibnamefont
  {Pedley}},\ }\href@noop {} {\bibfield  {journal} {\bibinfo  {journal}
  {Journal of Fluid Mechanics}\ }\textbf {\bibinfo {volume} {324}},\ \bibinfo
  {pages} {223} (\bibinfo {year} {1996})}\BibitemShut {NoStop}%
\bibitem [{\citenamefont {Kaya}\ and\ \citenamefont
  {Koser}(2012)}]{kaya2012direct}%
  \BibitemOpen
  \bibfield  {author} {\bibinfo {author} {\bibfnamefont {T.}~\bibnamefont
  {Kaya}}\ and\ \bibinfo {author} {\bibfnamefont {H.}~\bibnamefont {Koser}},\
  }\href@noop {} {\bibfield  {journal} {\bibinfo  {journal} {Biophysical
  journal}\ }\textbf {\bibinfo {volume} {102}},\ \bibinfo {pages} {1514}
  (\bibinfo {year} {2012})}\BibitemShut {NoStop}%
\bibitem [{\citenamefont {Kantsler}\ \emph {et~al.}(2014)\citenamefont
  {Kantsler}, \citenamefont {Dunkel}, \citenamefont {Blayney},\ and\
  \citenamefont {Goldstein}}]{kantsler2014rheotaxis}%
  \BibitemOpen
  \bibfield  {author} {\bibinfo {author} {\bibfnamefont {V.}~\bibnamefont
  {Kantsler}}, \bibinfo {author} {\bibfnamefont {J.}~\bibnamefont {Dunkel}},
  \bibinfo {author} {\bibfnamefont {M.}~\bibnamefont {Blayney}},\ and\ \bibinfo
  {author} {\bibfnamefont {R.~E.}\ \bibnamefont {Goldstein}},\ }\href@noop {}
  {\bibfield  {journal} {\bibinfo  {journal} {Elife}\ }\textbf {\bibinfo
  {volume} {3}},\ \bibinfo {pages} {e02403} (\bibinfo {year}
  {2014})}\BibitemShut {NoStop}%
\bibitem [{\citenamefont {Ishimoto}\ and\ \citenamefont
  {Gaffney}(2015)}]{ishimoto2015fluid}%
  \BibitemOpen
  \bibfield  {author} {\bibinfo {author} {\bibfnamefont {K.}~\bibnamefont
  {Ishimoto}}\ and\ \bibinfo {author} {\bibfnamefont {E.~A.}\ \bibnamefont
  {Gaffney}},\ }\href@noop {} {\bibfield  {journal} {\bibinfo  {journal}
  {Journal of The Royal Society Interface}\ }\textbf {\bibinfo {volume} {12}},\
  \bibinfo {pages} {20150172} (\bibinfo {year} {2015})}\BibitemShut {NoStop}%
\bibitem [{\citenamefont {Mathijssen}\ \emph {et~al.}(2019)\citenamefont
  {Mathijssen}, \citenamefont {Figueroa-Morales}, \citenamefont {Junot},
  \citenamefont {Cl{\'e}ment}, \citenamefont {Lindner},\ and\ \citenamefont
  {Z{\"o}ttl}}]{mathijssen2019oscillatory}%
  \BibitemOpen
  \bibfield  {author} {\bibinfo {author} {\bibfnamefont {A.~J.}\ \bibnamefont
  {Mathijssen}}, \bibinfo {author} {\bibfnamefont {N.}~\bibnamefont
  {Figueroa-Morales}}, \bibinfo {author} {\bibfnamefont {G.}~\bibnamefont
  {Junot}}, \bibinfo {author} {\bibfnamefont {{\'E}.}~\bibnamefont
  {Cl{\'e}ment}}, \bibinfo {author} {\bibfnamefont {A.}~\bibnamefont
  {Lindner}},\ and\ \bibinfo {author} {\bibfnamefont {A.}~\bibnamefont
  {Z{\"o}ttl}},\ }\href@noop {} {\bibfield  {journal} {\bibinfo  {journal}
  {Nature communications}\ }\textbf {\bibinfo {volume} {10}},\ \bibinfo {pages}
  {3434} (\bibinfo {year} {2019})}\BibitemShut {NoStop}%
\bibitem [{\citenamefont {Ishimoto}(2023)}]{ishimoto2023jeffery}%
  \BibitemOpen
  \bibfield  {author} {\bibinfo {author} {\bibfnamefont {K.}~\bibnamefont
  {Ishimoto}},\ }\href@noop {} {\bibfield  {journal} {\bibinfo  {journal}
  {Journal of the Physical Society of Japan}\ }\textbf {\bibinfo {volume}
  {92}},\ \bibinfo {pages} {062001} (\bibinfo {year} {2023})}\BibitemShut
  {NoStop}%
\bibitem [{\citenamefont {Soto}\ \emph {et~al.}(2022)\citenamefont {Soto},
  \citenamefont {Karshalev}, \citenamefont {Zhang}, \citenamefont {Esteban
  Fernandez~de Avila}, \citenamefont {Nourhani},\ and\ \citenamefont
  {Wang}}]{soto2022smart}%
  \BibitemOpen
  \bibfield  {author} {\bibinfo {author} {\bibfnamefont {F.}~\bibnamefont
  {Soto}}, \bibinfo {author} {\bibfnamefont {E.}~\bibnamefont {Karshalev}},
  \bibinfo {author} {\bibfnamefont {F.}~\bibnamefont {Zhang}}, \bibinfo
  {author} {\bibfnamefont {B.}~\bibnamefont {Esteban Fernandez~de Avila}},
  \bibinfo {author} {\bibfnamefont {A.}~\bibnamefont {Nourhani}},\ and\
  \bibinfo {author} {\bibfnamefont {J.}~\bibnamefont {Wang}},\ }\href@noop {}
  {\bibfield  {journal} {\bibinfo  {journal} {Chemical Reviews}\ }\textbf
  {\bibinfo {volume} {122}},\ \bibinfo {pages} {5365} (\bibinfo {year}
  {2022})}\BibitemShut {NoStop}%
\bibitem [{\citenamefont {Essafri}\ \emph {et~al.}(2022)\citenamefont
  {Essafri}, \citenamefont {Ghosh}, \citenamefont {Desgranges},\ and\
  \citenamefont {Delhommelle}}]{essafri2022designing}%
  \BibitemOpen
  \bibfield  {author} {\bibinfo {author} {\bibfnamefont {I.}~\bibnamefont
  {Essafri}}, \bibinfo {author} {\bibfnamefont {B.}~\bibnamefont {Ghosh}},
  \bibinfo {author} {\bibfnamefont {C.}~\bibnamefont {Desgranges}},\ and\
  \bibinfo {author} {\bibfnamefont {J.}~\bibnamefont {Delhommelle}},\
  }\href@noop {} {\bibfield  {journal} {\bibinfo  {journal} {Physics of
  Fluids}\ }\textbf {\bibinfo {volume} {34}} (\bibinfo {year}
  {2022})}\BibitemShut {NoStop}%
\bibitem [{\citenamefont {Tsang}\ \emph {et~al.}(2020)\citenamefont {Tsang},
  \citenamefont {Demir}, \citenamefont {Ding},\ and\ \citenamefont
  {Pak}}]{tsang2020roads}%
  \BibitemOpen
  \bibfield  {author} {\bibinfo {author} {\bibfnamefont {A.~C.}\ \bibnamefont
  {Tsang}}, \bibinfo {author} {\bibfnamefont {E.}~\bibnamefont {Demir}},
  \bibinfo {author} {\bibfnamefont {Y.}~\bibnamefont {Ding}},\ and\ \bibinfo
  {author} {\bibfnamefont {O.~S.}\ \bibnamefont {Pak}},\ }\href@noop {}
  {\bibfield  {journal} {\bibinfo  {journal} {Advanced Intelligent Systems}\
  }\textbf {\bibinfo {volume} {2}},\ \bibinfo {pages} {1900137} (\bibinfo
  {year} {2020})}\BibitemShut {NoStop}%
\bibitem [{\citenamefont {Stark}(2021)}]{stark2021artificial}%
  \BibitemOpen
  \bibfield  {author} {\bibinfo {author} {\bibfnamefont {H.}~\bibnamefont
  {Stark}},\ }\href@noop {} {\bibfield  {journal} {\bibinfo  {journal} {Science
  Robotics}\ }\textbf {\bibinfo {volume} {6}},\ \bibinfo {pages} {eabh1977}
  (\bibinfo {year} {2021})}\BibitemShut {NoStop}%
\bibitem [{\citenamefont {Takagi}\ \emph {et~al.}(2014)\citenamefont {Takagi},
  \citenamefont {Palacci}, \citenamefont {Braunschweig}, \citenamefont
  {Shelley},\ and\ \citenamefont {Zhang}}]{takagi2014hydrodynamic}%
  \BibitemOpen
  \bibfield  {author} {\bibinfo {author} {\bibfnamefont {D.}~\bibnamefont
  {Takagi}}, \bibinfo {author} {\bibfnamefont {J.}~\bibnamefont {Palacci}},
  \bibinfo {author} {\bibfnamefont {A.~B.}\ \bibnamefont {Braunschweig}},
  \bibinfo {author} {\bibfnamefont {M.~J.}\ \bibnamefont {Shelley}},\ and\
  \bibinfo {author} {\bibfnamefont {J.}~\bibnamefont {Zhang}},\ }\href@noop {}
  {\bibfield  {journal} {\bibinfo  {journal} {Soft Matter}\ }\textbf {\bibinfo
  {volume} {10}},\ \bibinfo {pages} {1784} (\bibinfo {year}
  {2014})}\BibitemShut {NoStop}%
\bibitem [{\citenamefont {Ishimoto}\ and\ \citenamefont
  {Gaffney}(2014)}]{ishimoto2014study}%
  \BibitemOpen
  \bibfield  {author} {\bibinfo {author} {\bibfnamefont {K.}~\bibnamefont
  {Ishimoto}}\ and\ \bibinfo {author} {\bibfnamefont {E.~A.}\ \bibnamefont
  {Gaffney}},\ }\href@noop {} {\bibfield  {journal} {\bibinfo  {journal}
  {Journal of Theoretical Biology}\ }\textbf {\bibinfo {volume} {360}},\
  \bibinfo {pages} {187} (\bibinfo {year} {2014})}\BibitemShut {NoStop}%
\bibitem [{\citenamefont {Shum}\ and\ \citenamefont
  {Gaffney}(2015)}]{shum2015hydrodynamic}%
  \BibitemOpen
  \bibfield  {author} {\bibinfo {author} {\bibfnamefont {H.}~\bibnamefont
  {Shum}}\ and\ \bibinfo {author} {\bibfnamefont {E.~A.}\ \bibnamefont
  {Gaffney}},\ }\href@noop {} {\bibfield  {journal} {\bibinfo  {journal}
  {Physical Review E}\ }\textbf {\bibinfo {volume} {91}},\ \bibinfo {pages}
  {033012} (\bibinfo {year} {2015})}\BibitemShut {NoStop}%
\bibitem [{\citenamefont {Chamolly}\ \emph {et~al.}(2017)\citenamefont
  {Chamolly}, \citenamefont {Ishikawa},\ and\ \citenamefont
  {Lauga}}]{chamolly2017active}%
  \BibitemOpen
  \bibfield  {author} {\bibinfo {author} {\bibfnamefont {A.}~\bibnamefont
  {Chamolly}}, \bibinfo {author} {\bibfnamefont {T.}~\bibnamefont {Ishikawa}},\
  and\ \bibinfo {author} {\bibfnamefont {E.}~\bibnamefont {Lauga}},\
  }\href@noop {} {\bibfield  {journal} {\bibinfo  {journal} {New Journal of
  Physics}\ }\textbf {\bibinfo {volume} {19}},\ \bibinfo {pages} {115001}
  (\bibinfo {year} {2017})}\BibitemShut {NoStop}%
\bibitem [{\citenamefont {Reigh}\ \emph {et~al.}(2017)\citenamefont {Reigh},
  \citenamefont {Zhu}, \citenamefont {Gallaire},\ and\ \citenamefont
  {Lauga}}]{reigh2017}%
  \BibitemOpen
  \bibfield  {author} {\bibinfo {author} {\bibfnamefont {S.~Y.}\ \bibnamefont
  {Reigh}}, \bibinfo {author} {\bibfnamefont {L.}~\bibnamefont {Zhu}}, \bibinfo
  {author} {\bibfnamefont {F.}~\bibnamefont {Gallaire}},\ and\ \bibinfo
  {author} {\bibfnamefont {E.}~\bibnamefont {Lauga}},\ }\href
  {https://doi.org/10.1039/C6SM01636G} {\bibfield  {journal} {\bibinfo
  {journal} {Soft Matter}\ }\textbf {\bibinfo {volume} {13}},\ \bibinfo {pages}
  {3161} (\bibinfo {year} {2017})}\BibitemShut {NoStop}%
\bibitem [{\citenamefont {Shaik}\ \emph {et~al.}(2018)\citenamefont {Shaik},
  \citenamefont {Vasani},\ and\ \citenamefont {Ardekani}}]{shaik2018}%
  \BibitemOpen
  \bibfield  {author} {\bibinfo {author} {\bibfnamefont {V.~A.}\ \bibnamefont
  {Shaik}}, \bibinfo {author} {\bibfnamefont {V.}~\bibnamefont {Vasani}},\ and\
  \bibinfo {author} {\bibfnamefont {A.~M.}\ \bibnamefont {Ardekani}},\ }\href
  {https://doi.org/10.1017/jfm.2018.491} {\bibfield  {journal} {\bibinfo
  {journal} {Journal of Fluid Mechanics}\ }\textbf {\bibinfo {volume} {851}},\
  \bibinfo {pages} {187} (\bibinfo {year} {2018})}\BibitemShut {NoStop}%
\bibitem [{\citenamefont {Huang}\ \emph {et~al.}(2020)\citenamefont {Huang},
  \citenamefont {Omori},\ and\ \citenamefont {Ishikawa}}]{huang2020}%
  \BibitemOpen
  \bibfield  {author} {\bibinfo {author} {\bibfnamefont {Z.}~\bibnamefont
  {Huang}}, \bibinfo {author} {\bibfnamefont {T.}~\bibnamefont {Omori}},\ and\
  \bibinfo {author} {\bibfnamefont {T.}~\bibnamefont {Ishikawa}},\ }\href
  {https://doi.org/10.1103/PhysRevE.102.022603} {\bibfield  {journal} {\bibinfo
   {journal} {Physical Review E}\ }\textbf {\bibinfo {volume} {102}},\ \bibinfo
  {pages} {022603} (\bibinfo {year} {2020})}\BibitemShut {NoStop}%
\bibitem [{\citenamefont {Sprenger}\ \emph {et~al.}(2020)\citenamefont
  {Sprenger}, \citenamefont {Shaik}, \citenamefont {Ardekani}, \citenamefont
  {Lisicki}, \citenamefont {Mathijssen}, \citenamefont {Guzmán-Lastra},
  \citenamefont {Löwen}, \citenamefont {Menzel},\ and\ \citenamefont
  {Daddi-Moussa-Ider}}]{sprenger2020}%
  \BibitemOpen
  \bibfield  {author} {\bibinfo {author} {\bibfnamefont {A.~R.}\ \bibnamefont
  {Sprenger}}, \bibinfo {author} {\bibfnamefont {V.~A.}\ \bibnamefont {Shaik}},
  \bibinfo {author} {\bibfnamefont {A.~M.}\ \bibnamefont {Ardekani}}, \bibinfo
  {author} {\bibfnamefont {M.}~\bibnamefont {Lisicki}}, \bibinfo {author}
  {\bibfnamefont {A.~J. T.~M.}\ \bibnamefont {Mathijssen}}, \bibinfo {author}
  {\bibfnamefont {F.}~\bibnamefont {Guzmán-Lastra}}, \bibinfo {author}
  {\bibfnamefont {H.}~\bibnamefont {Löwen}}, \bibinfo {author} {\bibfnamefont
  {A.~M.}\ \bibnamefont {Menzel}},\ and\ \bibinfo {author} {\bibfnamefont
  {A.}~\bibnamefont {Daddi-Moussa-Ider}},\ }\href
  {https://doi.org/10.1140/epje/i2020-11980-9} {\bibfield  {journal} {\bibinfo
  {journal} {The European Physical Journal E}\ }\textbf {\bibinfo {volume}
  {43}},\ \bibinfo {pages} {58} (\bibinfo {year} {2020})}\BibitemShut {NoStop}%
\bibitem [{\citenamefont {Purcell}(1977)}]{Purcell1977}%
  \BibitemOpen
  \bibfield  {author} {\bibinfo {author} {\bibfnamefont {E.~M.}\ \bibnamefont
  {Purcell}},\ }\href {https://doi.org/10.1119/1.10903} {\bibfield  {journal}
  {\bibinfo  {journal} {American Journal of Physics}\ }\textbf {\bibinfo
  {volume} {45}},\ \bibinfo {pages} {3} (\bibinfo {year} {1977})}\BibitemShut
  {NoStop}%
\bibitem [{\citenamefont {Shapere}\ and\ \citenamefont
  {Wilczek}(1989)}]{Shapere1989}%
  \BibitemOpen
  \bibfield  {author} {\bibinfo {author} {\bibfnamefont {A.}~\bibnamefont
  {Shapere}}\ and\ \bibinfo {author} {\bibfnamefont {F.}~\bibnamefont
  {Wilczek}},\ }\href {https://doi.org/10.1017/S002211208900025X} {\bibfield
  {journal} {\bibinfo  {journal} {Journal of Fluid Mechanics}\ }\textbf
  {\bibinfo {volume} {198}},\ \bibinfo {pages} {557} (\bibinfo {year}
  {1989})}\BibitemShut {NoStop}%
\bibitem [{\citenamefont {Lauga}(2011)}]{Lauga2011}%
  \BibitemOpen
  \bibfield  {author} {\bibinfo {author} {\bibfnamefont {E.}~\bibnamefont
  {Lauga}},\ }\href {https://doi.org/10.1039/c0sm00953a} {\bibfield  {journal}
  {\bibinfo  {journal} {Soft Matter}\ }\textbf {\bibinfo {volume} {7}},\
  \bibinfo {pages} {3060} (\bibinfo {year} {2011})}\BibitemShut {NoStop}%
\bibitem [{\citenamefont {Ishimoto}\ and\ \citenamefont
  {Yamada}(2012)}]{ishimoto2012coordinate}%
  \BibitemOpen
  \bibfield  {author} {\bibinfo {author} {\bibfnamefont {K.}~\bibnamefont
  {Ishimoto}}\ and\ \bibinfo {author} {\bibfnamefont {M.}~\bibnamefont
  {Yamada}},\ }\href@noop {} {\bibfield  {journal} {\bibinfo  {journal} {SIAM
  Journal on Applied Mathematics}\ }\textbf {\bibinfo {volume} {72}},\ \bibinfo
  {pages} {1686} (\bibinfo {year} {2012})}\BibitemShut {NoStop}%
\bibitem [{\citenamefont {Dunstan}\ \emph {et~al.}(2012)\citenamefont
  {Dunstan}, \citenamefont {Miño}, \citenamefont {Clement},\ and\
  \citenamefont {Soto}}]{dunstan2012}%
  \BibitemOpen
  \bibfield  {author} {\bibinfo {author} {\bibfnamefont {J.}~\bibnamefont
  {Dunstan}}, \bibinfo {author} {\bibfnamefont {G.}~\bibnamefont {Miño}},
  \bibinfo {author} {\bibfnamefont {E.}~\bibnamefont {Clement}},\ and\ \bibinfo
  {author} {\bibfnamefont {R.}~\bibnamefont {Soto}},\ }\href
  {https://doi.org/10.1063/1.3676245} {\bibfield  {journal} {\bibinfo
  {journal} {Physics of Fluids}\ }\textbf {\bibinfo {volume} {24}},\ \bibinfo
  {pages} {011901} (\bibinfo {year} {2012})}\BibitemShut {NoStop}%
\bibitem [{\citenamefont {Datt}\ \emph {et~al.}(2018)\citenamefont {Datt},
  \citenamefont {Nasouri},\ and\ \citenamefont {Elfring}}]{datt2018two}%
  \BibitemOpen
  \bibfield  {author} {\bibinfo {author} {\bibfnamefont {C.}~\bibnamefont
  {Datt}}, \bibinfo {author} {\bibfnamefont {B.}~\bibnamefont {Nasouri}},\ and\
  \bibinfo {author} {\bibfnamefont {G.~J.}\ \bibnamefont {Elfring}},\
  }\href@noop {} {\bibfield  {journal} {\bibinfo  {journal} {Physical Review
  Fluids}\ }\textbf {\bibinfo {volume} {3}},\ \bibinfo {pages} {123301}
  (\bibinfo {year} {2018})}\BibitemShut {NoStop}%
\bibitem [{\citenamefont {Eberhard}\ \emph {et~al.}(2023)\citenamefont
  {Eberhard}, \citenamefont {Choudhary},\ and\ \citenamefont
  {Stark}}]{eberhard2023}%
  \BibitemOpen
  \bibfield  {author} {\bibinfo {author} {\bibfnamefont {M.}~\bibnamefont
  {Eberhard}}, \bibinfo {author} {\bibfnamefont {A.}~\bibnamefont
  {Choudhary}},\ and\ \bibinfo {author} {\bibfnamefont {H.}~\bibnamefont
  {Stark}},\ }\href {https://doi.org/10.1063/5.0151585} {\bibfield  {journal}
  {\bibinfo  {journal} {Physics of Fluids}\ }\textbf {\bibinfo {volume} {35}},\
  \bibinfo {pages} {063119} (\bibinfo {year} {2023})}\BibitemShut {NoStop}%
\bibitem [{\citenamefont {Najafi}\ and\ \citenamefont
  {Golestanian}(2004)}]{najafi2004simple}%
  \BibitemOpen
  \bibfield  {author} {\bibinfo {author} {\bibfnamefont {A.}~\bibnamefont
  {Najafi}}\ and\ \bibinfo {author} {\bibfnamefont {R.}~\bibnamefont
  {Golestanian}},\ }\href@noop {} {\bibfield  {journal} {\bibinfo  {journal}
  {Physical Review E}\ }\textbf {\bibinfo {volume} {69}},\ \bibinfo {pages}
  {062901} (\bibinfo {year} {2004})}\BibitemShut {NoStop}%
\bibitem [{\citenamefont {Golestanian}\ and\ \citenamefont
  {Ajdari}(2008{\natexlab{a}})}]{golestanian2008analytic}%
  \BibitemOpen
  \bibfield  {author} {\bibinfo {author} {\bibfnamefont {R.}~\bibnamefont
  {Golestanian}}\ and\ \bibinfo {author} {\bibfnamefont {A.}~\bibnamefont
  {Ajdari}},\ }\href@noop {} {\bibfield  {journal} {\bibinfo  {journal}
  {Physical Review E}\ }\textbf {\bibinfo {volume} {77}},\ \bibinfo {pages}
  {036308} (\bibinfo {year} {2008}{\natexlab{a}})}\BibitemShut {NoStop}%
\bibitem [{\citenamefont {Yasuda}\ \emph {et~al.}(2023)\citenamefont {Yasuda},
  \citenamefont {Hosaka},\ and\ \citenamefont
  {Komura}}]{yasuda2023generalized}%
  \BibitemOpen
  \bibfield  {author} {\bibinfo {author} {\bibfnamefont {K.}~\bibnamefont
  {Yasuda}}, \bibinfo {author} {\bibfnamefont {Y.}~\bibnamefont {Hosaka}},\
  and\ \bibinfo {author} {\bibfnamefont {S.}~\bibnamefont {Komura}},\
  }\href@noop {} {\bibfield  {journal} {\bibinfo  {journal} {arXiv preprint
  arXiv:2305.08411}\ } (\bibinfo {year} {2023})}\BibitemShut {NoStop}%
\bibitem [{\citenamefont {Lauga}(2007)}]{lauga2007continuous}%
  \BibitemOpen
  \bibfield  {author} {\bibinfo {author} {\bibfnamefont {E.}~\bibnamefont
  {Lauga}},\ }\href@noop {} {\bibfield  {journal} {\bibinfo  {journal} {Physics
  of Fluids}\ }\textbf {\bibinfo {volume} {19}},\ \bibinfo {pages} {061703}
  (\bibinfo {year} {2007})}\BibitemShut {NoStop}%
\bibitem [{\citenamefont {Ishimoto}(2013)}]{ishimoto2013spherical}%
  \BibitemOpen
  \bibfield  {author} {\bibinfo {author} {\bibfnamefont {K.}~\bibnamefont
  {Ishimoto}},\ }\href@noop {} {\bibfield  {journal} {\bibinfo  {journal}
  {Journal of Fluid Mechanics}\ }\textbf {\bibinfo {volume} {723}},\ \bibinfo
  {pages} {163} (\bibinfo {year} {2013})}\BibitemShut {NoStop}%
\bibitem [{\citenamefont {Hubert}\ \emph {et~al.}(2021)\citenamefont {Hubert},
  \citenamefont {Trosman}, \citenamefont {Collard}, \citenamefont {Sukhov},
  \citenamefont {Harting}, \citenamefont {Vandewalle},\ and\ \citenamefont
  {Smith}}]{hubert2021scallop}%
  \BibitemOpen
  \bibfield  {author} {\bibinfo {author} {\bibfnamefont {M.}~\bibnamefont
  {Hubert}}, \bibinfo {author} {\bibfnamefont {O.}~\bibnamefont {Trosman}},
  \bibinfo {author} {\bibfnamefont {Y.}~\bibnamefont {Collard}}, \bibinfo
  {author} {\bibfnamefont {A.}~\bibnamefont {Sukhov}}, \bibinfo {author}
  {\bibfnamefont {J.}~\bibnamefont {Harting}}, \bibinfo {author} {\bibfnamefont
  {N.}~\bibnamefont {Vandewalle}},\ and\ \bibinfo {author} {\bibfnamefont
  {A.-S.}\ \bibnamefont {Smith}},\ }\href@noop {} {\bibfield  {journal}
  {\bibinfo  {journal} {Physical review letters}\ }\textbf {\bibinfo {volume}
  {126}},\ \bibinfo {pages} {224501} (\bibinfo {year} {2021})}\BibitemShut
  {NoStop}%
\bibitem [{\citenamefont {Derr}\ \emph {et~al.}(2022)\citenamefont {Derr},
  \citenamefont {Dombrowski}, \citenamefont {Rycroft},\ and\ \citenamefont
  {Klotsa}}]{derr2022reciprocal}%
  \BibitemOpen
  \bibfield  {author} {\bibinfo {author} {\bibfnamefont {N.~J.}\ \bibnamefont
  {Derr}}, \bibinfo {author} {\bibfnamefont {T.}~\bibnamefont {Dombrowski}},
  \bibinfo {author} {\bibfnamefont {C.~H.}\ \bibnamefont {Rycroft}},\ and\
  \bibinfo {author} {\bibfnamefont {D.}~\bibnamefont {Klotsa}},\ }\href@noop {}
  {\bibfield  {journal} {\bibinfo  {journal} {Journal of Fluid Mechanics}\
  }\textbf {\bibinfo {volume} {952}},\ \bibinfo {pages} {A8} (\bibinfo {year}
  {2022})}\BibitemShut {NoStop}%
\bibitem [{\citenamefont {Curtis}\ and\ \citenamefont
  {Gaffney}(2013)}]{Curtis2013}%
  \BibitemOpen
  \bibfield  {author} {\bibinfo {author} {\bibfnamefont {M.~P.}\ \bibnamefont
  {Curtis}}\ and\ \bibinfo {author} {\bibfnamefont {E.~A.}\ \bibnamefont
  {Gaffney}},\ }\bibfield  {journal} {\bibinfo  {journal} {Physical Review E -
  Statistical, Nonlinear, and Soft Matter Physics}\ }\textbf {\bibinfo {volume}
  {87}},\ \href {https://doi.org/10.1103/PhysRevE.87.043006}
  {10.1103/PhysRevE.87.043006} (\bibinfo {year} {2013})\BibitemShut {NoStop}%
\bibitem [{\citenamefont {Lauga}(2014)}]{lauga2014locomotion}%
  \BibitemOpen
  \bibfield  {author} {\bibinfo {author} {\bibfnamefont {E.}~\bibnamefont
  {Lauga}},\ }\href@noop {} {\bibfield  {journal} {\bibinfo  {journal} {Physics
  of Fluids}\ }\textbf {\bibinfo {volume} {26}},\ \bibinfo {pages} {081902}
  (\bibinfo {year} {2014})}\BibitemShut {NoStop}%
\bibitem [{\citenamefont {Qiu}\ \emph {et~al.}(2014)\citenamefont {Qiu},
  \citenamefont {Lee}, \citenamefont {Mark}, \citenamefont {Morozov},
  \citenamefont {M{\"u}nster}, \citenamefont {Mierka}, \citenamefont {Turek},
  \citenamefont {Leshansky},\ and\ \citenamefont {Fischer}}]{qiu2014swimming}%
  \BibitemOpen
  \bibfield  {author} {\bibinfo {author} {\bibfnamefont {T.}~\bibnamefont
  {Qiu}}, \bibinfo {author} {\bibfnamefont {T.-C.}\ \bibnamefont {Lee}},
  \bibinfo {author} {\bibfnamefont {A.~G.}\ \bibnamefont {Mark}}, \bibinfo
  {author} {\bibfnamefont {K.~I.}\ \bibnamefont {Morozov}}, \bibinfo {author}
  {\bibfnamefont {R.}~\bibnamefont {M{\"u}nster}}, \bibinfo {author}
  {\bibfnamefont {O.}~\bibnamefont {Mierka}}, \bibinfo {author} {\bibfnamefont
  {S.}~\bibnamefont {Turek}}, \bibinfo {author} {\bibfnamefont {A.~M.}\
  \bibnamefont {Leshansky}},\ and\ \bibinfo {author} {\bibfnamefont
  {P.}~\bibnamefont {Fischer}},\ }\href@noop {} {\bibfield  {journal} {\bibinfo
   {journal} {Nature communications}\ }\textbf {\bibinfo {volume} {5}},\
  \bibinfo {pages} {5119} (\bibinfo {year} {2014})}\BibitemShut {NoStop}%
\bibitem [{\citenamefont {Yasuda}\ \emph {et~al.}(2020)\citenamefont {Yasuda},
  \citenamefont {Kuroda},\ and\ \citenamefont {Komura}}]{yasuda2020reciprocal}%
  \BibitemOpen
  \bibfield  {author} {\bibinfo {author} {\bibfnamefont {K.}~\bibnamefont
  {Yasuda}}, \bibinfo {author} {\bibfnamefont {M.}~\bibnamefont {Kuroda}},\
  and\ \bibinfo {author} {\bibfnamefont {S.}~\bibnamefont {Komura}},\
  }\href@noop {} {\bibfield  {journal} {\bibinfo  {journal} {Physics of
  Fluids}\ }\textbf {\bibinfo {volume} {32}},\ \bibinfo {pages} {093102}
  (\bibinfo {year} {2020})}\BibitemShut {NoStop}%
\bibitem [{\citenamefont {Romanczuk}\ \emph {et~al.}(2012)\citenamefont
  {Romanczuk}, \citenamefont {B{\"a}r}, \citenamefont {Ebeling}, \citenamefont
  {Lindner},\ and\ \citenamefont {Schimansky-Geier}}]{romanczuk2012active}%
  \BibitemOpen
  \bibfield  {author} {\bibinfo {author} {\bibfnamefont {P.}~\bibnamefont
  {Romanczuk}}, \bibinfo {author} {\bibfnamefont {M.}~\bibnamefont {B{\"a}r}},
  \bibinfo {author} {\bibfnamefont {W.}~\bibnamefont {Ebeling}}, \bibinfo
  {author} {\bibfnamefont {B.}~\bibnamefont {Lindner}},\ and\ \bibinfo {author}
  {\bibfnamefont {L.}~\bibnamefont {Schimansky-Geier}},\ }\href@noop {}
  {\bibfield  {journal} {\bibinfo  {journal} {The European Physical Journal
  Special Topics}\ }\textbf {\bibinfo {volume} {202}},\ \bibinfo {pages} {1}
  (\bibinfo {year} {2012})}\BibitemShut {NoStop}%
\bibitem [{\citenamefont {Bechinger}\ \emph {et~al.}(2016)\citenamefont
  {Bechinger}, \citenamefont {Leonardo}, \citenamefont {Löwen}, \citenamefont
  {Reichhardt}, \citenamefont {Volpe},\ and\ \citenamefont
  {Volpe}}]{Bechinger2016}%
  \BibitemOpen
  \bibfield  {author} {\bibinfo {author} {\bibfnamefont {C.}~\bibnamefont
  {Bechinger}}, \bibinfo {author} {\bibfnamefont {R.~D.}\ \bibnamefont
  {Leonardo}}, \bibinfo {author} {\bibfnamefont {H.}~\bibnamefont {Löwen}},
  \bibinfo {author} {\bibfnamefont {C.}~\bibnamefont {Reichhardt}}, \bibinfo
  {author} {\bibfnamefont {G.}~\bibnamefont {Volpe}},\ and\ \bibinfo {author}
  {\bibfnamefont {G.}~\bibnamefont {Volpe}},\ }\bibfield  {journal} {\bibinfo
  {journal} {Reviews of Modern Physics}\ }\textbf {\bibinfo {volume} {88}},\
  \href {https://doi.org/10.1103/RevModPhys.88.045006}
  {10.1103/RevModPhys.88.045006} (\bibinfo {year} {2016})\BibitemShut {NoStop}%
\bibitem [{\citenamefont {Patch}\ \emph {et~al.}(2017)\citenamefont {Patch},
  \citenamefont {Yllanes},\ and\ \citenamefont
  {Marchetti}}]{patch2017kinetics}%
  \BibitemOpen
  \bibfield  {author} {\bibinfo {author} {\bibfnamefont {A.}~\bibnamefont
  {Patch}}, \bibinfo {author} {\bibfnamefont {D.}~\bibnamefont {Yllanes}},\
  and\ \bibinfo {author} {\bibfnamefont {M.~C.}\ \bibnamefont {Marchetti}},\
  }\href@noop {} {\bibfield  {journal} {\bibinfo  {journal} {Physical Review
  E}\ }\textbf {\bibinfo {volume} {95}},\ \bibinfo {pages} {012601} (\bibinfo
  {year} {2017})}\BibitemShut {NoStop}%
\bibitem [{\citenamefont {B{\"a}r}\ \emph {et~al.}(2020)\citenamefont
  {B{\"a}r}, \citenamefont {Gro{\ss}mann}, \citenamefont {Heidenreich},\ and\
  \citenamefont {Peruani}}]{bar2020self}%
  \BibitemOpen
  \bibfield  {author} {\bibinfo {author} {\bibfnamefont {M.}~\bibnamefont
  {B{\"a}r}}, \bibinfo {author} {\bibfnamefont {R.}~\bibnamefont
  {Gro{\ss}mann}}, \bibinfo {author} {\bibfnamefont {S.}~\bibnamefont
  {Heidenreich}},\ and\ \bibinfo {author} {\bibfnamefont {F.}~\bibnamefont
  {Peruani}},\ }\href@noop {} {\bibfield  {journal} {\bibinfo  {journal}
  {Annual Review of Condensed Matter Physics}\ }\textbf {\bibinfo {volume}
  {11}},\ \bibinfo {pages} {441} (\bibinfo {year} {2020})}\BibitemShut
  {NoStop}%
\bibitem [{\citenamefont {Yasuda}\ and\ \citenamefont
  {Ishimoto}(2022)}]{yasuda2022most}%
  \BibitemOpen
  \bibfield  {author} {\bibinfo {author} {\bibfnamefont {K.}~\bibnamefont
  {Yasuda}}\ and\ \bibinfo {author} {\bibfnamefont {K.}~\bibnamefont
  {Ishimoto}},\ }\href@noop {} {\bibfield  {journal} {\bibinfo  {journal}
  {Physical Review E}\ }\textbf {\bibinfo {volume} {106}},\ \bibinfo {pages}
  {064120} (\bibinfo {year} {2022})}\BibitemShut {NoStop}%
\bibitem [{\citenamefont {Golestanian}\ and\ \citenamefont
  {Ajdari}(2008{\natexlab{b}})}]{golestanian2008mechanical}%
  \BibitemOpen
  \bibfield  {author} {\bibinfo {author} {\bibfnamefont {R.}~\bibnamefont
  {Golestanian}}\ and\ \bibinfo {author} {\bibfnamefont {A.}~\bibnamefont
  {Ajdari}},\ }\href@noop {} {\bibfield  {journal} {\bibinfo  {journal}
  {Physical review letters}\ }\textbf {\bibinfo {volume} {100}},\ \bibinfo
  {pages} {038101} (\bibinfo {year} {2008}{\natexlab{b}})}\BibitemShut
  {NoStop}%
\bibitem [{\citenamefont {Yasuda}\ \emph {et~al.}(2021)\citenamefont {Yasuda},
  \citenamefont {Hosaka}, \citenamefont {Sou},\ and\ \citenamefont
  {Komura}}]{yasuda2021odd}%
  \BibitemOpen
  \bibfield  {author} {\bibinfo {author} {\bibfnamefont {K.}~\bibnamefont
  {Yasuda}}, \bibinfo {author} {\bibfnamefont {Y.}~\bibnamefont {Hosaka}},
  \bibinfo {author} {\bibfnamefont {I.}~\bibnamefont {Sou}},\ and\ \bibinfo
  {author} {\bibfnamefont {S.}~\bibnamefont {Komura}},\ }\href@noop {}
  {\bibfield  {journal} {\bibinfo  {journal} {Journal of the Physical Society
  of Japan}\ }\textbf {\bibinfo {volume} {90}},\ \bibinfo {pages} {075001}
  (\bibinfo {year} {2021})}\BibitemShut {NoStop}%
\bibitem [{\citenamefont {Ishimoto}\ \emph {et~al.}(2022)\citenamefont
  {Ishimoto}, \citenamefont {Moreau},\ and\ \citenamefont
  {Yasuda}}]{ishimoto2022self}%
  \BibitemOpen
  \bibfield  {author} {\bibinfo {author} {\bibfnamefont {K.}~\bibnamefont
  {Ishimoto}}, \bibinfo {author} {\bibfnamefont {C.}~\bibnamefont {Moreau}},\
  and\ \bibinfo {author} {\bibfnamefont {K.}~\bibnamefont {Yasuda}},\
  }\href@noop {} {\bibfield  {journal} {\bibinfo  {journal} {Physical Review
  E}\ }\textbf {\bibinfo {volume} {105}},\ \bibinfo {pages} {064603} (\bibinfo
  {year} {2022})}\BibitemShut {NoStop}%
\bibitem [{\citenamefont {Ishimoto}\ \emph {et~al.}(2023)\citenamefont
  {Ishimoto}, \citenamefont {Moreau},\ and\ \citenamefont
  {Yasuda}}]{ishimoto2023odd}%
  \BibitemOpen
  \bibfield  {author} {\bibinfo {author} {\bibfnamefont {K.}~\bibnamefont
  {Ishimoto}}, \bibinfo {author} {\bibfnamefont {C.}~\bibnamefont {Moreau}},\
  and\ \bibinfo {author} {\bibfnamefont {K.}~\bibnamefont {Yasuda}},\
  }\href@noop {} {\bibfield  {journal} {\bibinfo  {journal} {PRX Life}\
  }\textbf {\bibinfo {volume} {1}},\ \bibinfo {pages} {023002} (\bibinfo {year}
  {2023})}\BibitemShut {NoStop}%
\bibitem [{\citenamefont {Matse}\ \emph {et~al.}(2017)\citenamefont {Matse},
  \citenamefont {Chubynsky},\ and\ \citenamefont {Bechhoefer}}]{matse2017test}%
  \BibitemOpen
  \bibfield  {author} {\bibinfo {author} {\bibfnamefont {M.}~\bibnamefont
  {Matse}}, \bibinfo {author} {\bibfnamefont {M.~V.}\ \bibnamefont
  {Chubynsky}},\ and\ \bibinfo {author} {\bibfnamefont {J.}~\bibnamefont
  {Bechhoefer}},\ }\href@noop {} {\bibfield  {journal} {\bibinfo  {journal}
  {Physical Review E}\ }\textbf {\bibinfo {volume} {96}},\ \bibinfo {pages}
  {042604} (\bibinfo {year} {2017})}\BibitemShut {NoStop}%
\bibitem [{\citenamefont {Hosaka}\ \emph {et~al.}(2017)\citenamefont {Hosaka},
  \citenamefont {Yasuda}, \citenamefont {Sou}, \citenamefont {Okamoto},\ and\
  \citenamefont {Komura}}]{hosaka2017thermally}%
  \BibitemOpen
  \bibfield  {author} {\bibinfo {author} {\bibfnamefont {Y.}~\bibnamefont
  {Hosaka}}, \bibinfo {author} {\bibfnamefont {K.}~\bibnamefont {Yasuda}},
  \bibinfo {author} {\bibfnamefont {I.}~\bibnamefont {Sou}}, \bibinfo {author}
  {\bibfnamefont {R.}~\bibnamefont {Okamoto}},\ and\ \bibinfo {author}
  {\bibfnamefont {S.}~\bibnamefont {Komura}},\ }\href@noop {} {\bibfield
  {journal} {\bibinfo  {journal} {Journal of the Physical Society of Japan}\
  }\textbf {\bibinfo {volume} {86}},\ \bibinfo {pages} {113801} (\bibinfo
  {year} {2017})}\BibitemShut {NoStop}%
\bibitem [{\citenamefont {Lauga}\ and\ \citenamefont
  {Powers}(2009)}]{lauga2009hydrodynamics}%
  \BibitemOpen
  \bibfield  {author} {\bibinfo {author} {\bibfnamefont {E.}~\bibnamefont
  {Lauga}}\ and\ \bibinfo {author} {\bibfnamefont {T.~R.}\ \bibnamefont
  {Powers}},\ }\href@noop {} {\bibfield  {journal} {\bibinfo  {journal}
  {Reports on progress in physics}\ }\textbf {\bibinfo {volume} {72}},\
  \bibinfo {pages} {096601} (\bibinfo {year} {2009})}\BibitemShut {NoStop}%
\bibitem [{\citenamefont {Yariv}(2006)}]{yariv2006self}%
  \BibitemOpen
  \bibfield  {author} {\bibinfo {author} {\bibfnamefont {E.}~\bibnamefont
  {Yariv}},\ }\href@noop {} {\bibfield  {journal} {\bibinfo  {journal} {Journal
  of Fluid Mechanics}\ }\textbf {\bibinfo {volume} {550}},\ \bibinfo {pages}
  {139} (\bibinfo {year} {2006})}\BibitemShut {NoStop}%
\bibitem [{\citenamefont {Sancho}\ \emph {et~al.}(1982)\citenamefont {Sancho},
  \citenamefont {Miguel},\ and\ \citenamefont {Dürr}}]{Sancho1982}%
  \BibitemOpen
  \bibfield  {author} {\bibinfo {author} {\bibfnamefont {J.~M.}\ \bibnamefont
  {Sancho}}, \bibinfo {author} {\bibfnamefont {M.~S.}\ \bibnamefont {Miguel}},\
  and\ \bibinfo {author} {\bibfnamefont {D.}~\bibnamefont {Dürr}},\ }\href
  {https://doi.org/10.1007/BF01012607} {\bibfield  {journal} {\bibinfo
  {journal} {Journal of Statistical Physics}\ }\textbf {\bibinfo {volume}
  {28}},\ \bibinfo {pages} {291} (\bibinfo {year} {1982})}\BibitemShut
  {NoStop}%
\bibitem [{\citenamefont {Pozrikidis}(1992)}]{Pozrikidis1992}%
  \BibitemOpen
  \bibfield  {author} {\bibinfo {author} {\bibfnamefont {C.}~\bibnamefont
  {Pozrikidis}},\ }\href {https://doi.org/10.1017/CBO9780511624124} {\emph
  {\bibinfo {title} {Boundary Integral and Singularity Methods for Linearized
  Viscous Flow}}}\ (\bibinfo  {publisher} {Cambridge University Press},\
  \bibinfo {year} {1992})\BibitemShut {NoStop}%
\bibitem [{\citenamefont {Maul}\ and\ \citenamefont
  {Kim}(1994)}]{maul1994image}%
  \BibitemOpen
  \bibfield  {author} {\bibinfo {author} {\bibfnamefont {C.}~\bibnamefont
  {Maul}}\ and\ \bibinfo {author} {\bibfnamefont {S.}~\bibnamefont {Kim}},\
  }\href@noop {} {\bibfield  {journal} {\bibinfo  {journal} {Physics of
  Fluids}\ }\textbf {\bibinfo {volume} {6}},\ \bibinfo {pages} {2221} (\bibinfo
  {year} {1994})}\BibitemShut {NoStop}%
\bibitem [{\citenamefont {Kim}\ and\ \citenamefont
  {Karrila}(2013)}]{kim2013microhydrodynamics}%
  \BibitemOpen
  \bibfield  {author} {\bibinfo {author} {\bibfnamefont {S.}~\bibnamefont
  {Kim}}\ and\ \bibinfo {author} {\bibfnamefont {S.~J.}\ \bibnamefont
  {Karrila}},\ }\href@noop {} {\emph {\bibinfo {title} {Microhydrodynamics:
  principles and selected applications}}}\ (\bibinfo  {publisher} {Courier
  Corporation},\ \bibinfo {year} {2013})\BibitemShut {NoStop}%
\bibitem [{\citenamefont {Maruyama}(1955)}]{Maruyama1955}%
  \BibitemOpen
  \bibfield  {author} {\bibinfo {author} {\bibfnamefont {G.}~\bibnamefont
  {Maruyama}},\ }\href {https://doi.org/10.1007/BF02846028} {\bibfield
  {journal} {\bibinfo  {journal} {Rendiconti del Circolo Matematico di
  Palermo}\ }\textbf {\bibinfo {volume} {4}},\ \bibinfo {pages} {48} (\bibinfo
  {year} {1955})}\BibitemShut {NoStop}%
\bibitem [{\citenamefont {Kloeden}\ and\ \citenamefont
  {Platen}(1992)}]{Kloeden1992}%
  \BibitemOpen
  \bibfield  {author} {\bibinfo {author} {\bibfnamefont {P.~E.}\ \bibnamefont
  {Kloeden}}\ and\ \bibinfo {author} {\bibfnamefont {E.}~\bibnamefont
  {Platen}},\ }\href {https://doi.org/10.1007/978-3-662-12616-5} {\emph
  {\bibinfo {title} {Numerical Solution of Stochastic Differential
  Equations}}}\ (\bibinfo  {publisher} {Springer Berlin Heidelberg},\ \bibinfo
  {year} {1992})\BibitemShut {NoStop}%
\bibitem [{\citenamefont {Kraichnan}(1959)}]{Kraichnan1959}%
  \BibitemOpen
  \bibfield  {author} {\bibinfo {author} {\bibfnamefont {R.~H.}\ \bibnamefont
  {Kraichnan}},\ }\href {https://doi.org/10.1017/S0022112059000362} {\bibfield
  {journal} {\bibinfo  {journal} {Journal of Fluid Mechanics}\ }\textbf
  {\bibinfo {volume} {5}},\ \bibinfo {pages} {497} (\bibinfo {year}
  {1959})}\BibitemShut {NoStop}%
\bibitem [{\citenamefont {Kida}\ and\ \citenamefont
  {Goto}(1997)}]{kida1997lagrangian}%
  \BibitemOpen
  \bibfield  {author} {\bibinfo {author} {\bibfnamefont {S.}~\bibnamefont
  {Kida}}\ and\ \bibinfo {author} {\bibfnamefont {S.}~\bibnamefont {Goto}},\
  }\href@noop {} {\bibfield  {journal} {\bibinfo  {journal} {Journal of Fluid
  Mechanics}\ }\textbf {\bibinfo {volume} {345}},\ \bibinfo {pages} {307}
  (\bibinfo {year} {1997})}\BibitemShut {NoStop}%
\bibitem [{\citenamefont {Goto}\ and\ \citenamefont {Kida}(2002)}]{Goto2002a}%
  \BibitemOpen
  \bibfield  {author} {\bibinfo {author} {\bibfnamefont {S.}~\bibnamefont
  {Goto}}\ and\ \bibinfo {author} {\bibfnamefont {S.}~\bibnamefont {Kida}},\
  }\href {https://doi.org/10.1088/0951-7715/15/5/309} {\bibfield  {journal}
  {\bibinfo  {journal} {Nonlinearity}\ }\textbf {\bibinfo {volume} {15}},\
  \bibinfo {pages} {309} (\bibinfo {year} {2002})}\BibitemShut {NoStop}%
\bibitem [{\citenamefont {Pontryagin}(2014)}]{pontryagin2014ordinary}%
  \BibitemOpen
  \bibfield  {author} {\bibinfo {author} {\bibfnamefont {L.~S.}\ \bibnamefont
  {Pontryagin}},\ }\href@noop {} {\emph {\bibinfo {title} {Ordinary
  Differential Equations: Adiwes International Series in Mathematics}}}\
  (\bibinfo  {publisher} {Elsevier},\ \bibinfo {year} {2014})\BibitemShut
  {NoStop}%
\bibitem [{\citenamefont {Ishimoto}(2019)}]{ishimoto2019bacterial}%
  \BibitemOpen
  \bibfield  {author} {\bibinfo {author} {\bibfnamefont {K.}~\bibnamefont
  {Ishimoto}},\ }\href@noop {} {\bibfield  {journal} {\bibinfo  {journal}
  {Journal of Fluid Mechanics}\ }\textbf {\bibinfo {volume} {880}},\ \bibinfo
  {pages} {620} (\bibinfo {year} {2019})}\BibitemShut {NoStop}%
\bibitem [{\citenamefont {Gnesotto}\ \emph {et~al.}(2018)\citenamefont
  {Gnesotto}, \citenamefont {Mura}, \citenamefont {Gladrow},\ and\
  \citenamefont {Broedersz}}]{gnesotto2018broken}%
  \BibitemOpen
  \bibfield  {author} {\bibinfo {author} {\bibfnamefont {F.~S.}\ \bibnamefont
  {Gnesotto}}, \bibinfo {author} {\bibfnamefont {F.}~\bibnamefont {Mura}},
  \bibinfo {author} {\bibfnamefont {J.}~\bibnamefont {Gladrow}},\ and\ \bibinfo
  {author} {\bibfnamefont {C.~P.}\ \bibnamefont {Broedersz}},\ }\href@noop {}
  {\bibfield  {journal} {\bibinfo  {journal} {Reports on Progress in Physics}\
  }\textbf {\bibinfo {volume} {81}},\ \bibinfo {pages} {066601} (\bibinfo
  {year} {2018})}\BibitemShut {NoStop}%
\bibitem [{\citenamefont {Ishikawa}\ \emph {et~al.}(2006)\citenamefont
  {Ishikawa}, \citenamefont {Simmonds},\ and\ \citenamefont
  {Pedley}}]{ishikawa2006hydrodynamic}%
  \BibitemOpen
  \bibfield  {author} {\bibinfo {author} {\bibfnamefont {T.}~\bibnamefont
  {Ishikawa}}, \bibinfo {author} {\bibfnamefont {M.}~\bibnamefont {Simmonds}},\
  and\ \bibinfo {author} {\bibfnamefont {T.~J.}\ \bibnamefont {Pedley}},\
  }\href@noop {} {\bibfield  {journal} {\bibinfo  {journal} {Journal of Fluid
  Mechanics}\ }\textbf {\bibinfo {volume} {568}},\ \bibinfo {pages} {119}
  (\bibinfo {year} {2006})}\BibitemShut {NoStop}%
\end{thebibliography}%

\end{document}